\pdfoutput=1
\documentclass[12pt]{article}
\usepackage{jheppub}

\usepackage[utf8]{inputenc}

\usepackage{amsmath}
\usepackage{amssymb}
\usepackage{graphicx,color}
\usepackage{amsthm}
\usepackage{mathrsfs}

\usepackage{tikz}
\usetikzlibrary{matrix,arrows}

\usepackage{ifpdf}

\usepackage{subfigure}

\usepackage{tikz}
\usetikzlibrary{shapes,shadows,calc}
\usepgflibrary{arrows}
\usepackage{rotating}
\usepackage{wrapfig}

\tikzset{
  sshadow/.style={opacity=.25, shadow xshift=0.05, shadow yshift=-0.06},
}

\def\kbbox[#1,#2,#3,#4,#5]#6{
        \draw[dashed] node[draw,color=gray!50,minimum
        height=#1,minimum width=#2] (#4) at #5 {}; 
        \node[anchor=#3,inner sep=2pt] at (#4.#3)  {#6};
}
\def\kbboxred[#1,#2,#3,#4,#5]#6{
        \draw[] node[draw,color=red,minimum
        height=#1,minimum width=#2] (#4) at #5 {}; 
        \node[anchor=#3,inner sep=2pt] at (#4.#3)  {#6};
}

\newcommand{\bP}{\mathbb{P}}

\newcommand{\bZ}{\mathbb{Z}}
\newcommand{\cA}{\mathcal{A}}

\newcommand{\cE}{\mathscr{E}}

\newcommand{\cI}{\mathcal{I}}
\newcommand{\cK}{\mathcal{K}}
\newcommand{\cL}{\mathcal{L}}

\newcommand{\cN}{\mathcal{N}}
\newcommand{\cO}{\mathcal{O}}

\newcommand{\cS}{\mathcal{S}}

\newcommand{\ov}{\overline}

\newcommand{\wt}[1]{\widetilde{#1}}

\newcommand{\dsz}[2]{\bigl\langle#1,#2\bigr\rangle}

\newcommand{\Kint}{\cK}

\DeclareMathOperator{\ch}{ch}
\DeclareMathOperator{\Td}{Td}

\DeclareMathOperator{\Res}{Res}

\def\Label#1{\label{#1}%
  \smash{\hbox to0pt{\raise1ex\hbox{\tiny[#1]}\hss}}}
\def\noLabels{\let\Label=\label}
\def\nobbibitem{\let\bbibitem=\bibitem}
 \def\noBibitem{\let\Bibitem=\bibitem}

\newcommand{\beq}{\begin{equation}}
\newcommand{\eeq}{\end{equation}}
\newcommand{\bea}{\begin{eqnarray}}
\newcommand{\eea}{\end{eqnarray}}

\newcommand{\I}{\text{Im}}

\newcommand{\nn}{\nonumber}

\renewcommand{\d}{\textrm{d}}

\newcommand\varpm{\mathbin{\vcenter{\hbox{%
  \oalign{\hfil$\scriptstyle+$\hfil\cr
          \noalign{\kern-.3ex}
          $\scriptscriptstyle({-})$\cr}%
}}}}
\newcommand\varmp{\mathbin{\vcenter{\hbox{%
  \oalign{$\scriptstyle({+})$\cr
          \noalign{\kern-.3ex}
          \hfil$\scriptscriptstyle-$\hfil\cr}%
}}}}

\title{\centering Special Points of Inflation in\\ 
 Flux Compactifications}

\author[a]{I\~naki Garc\'ia-Etxebarria,}
\author[a]{Thomas W. Grimm,}
\author[b]{and Irene Valenzuela}

\affiliation[a]{Max Planck Institute for Physics,\\
F\"ohringer Ring 6, 80805 Munich, Germany}

\affiliation[b]{Departamento de F\'{\i}sica Te\'orica 
and Instituto de F\'{\i}sica Te\'orica  UAM-CSIC, Universidad Aut\'onoma de Madrid,
Cantoblanco, 28049 Madrid, Spain}

\emailAdd{inaki@mpp.mpg.de}
\emailAdd{grimm@mpp.mpg.de}
\emailAdd{irene.valenzuela@uam.es}

\abstract{We study the realisation of axion inflation models in the
  complex structure moduli spaces of Calabi-Yau threefolds and
  fourfolds. The axions arise close to special points of these moduli
  spaces that admit discrete monodromy symmetries of infinite
  order. Examples include the large complex structure point and
  conifold point, but can be of more general nature. In Type IIB and
  F-theory compactifications the geometric axions receive a scalar
  potential from a flux-induced superpotential. We find toy variants
  of various inflationary potentials including the ones for natural
  inflation of one or multiple axions, or axion monodromy inflation
  with polynomial potential.  Interesting examples are also given by
  mirror geometries of torus fibrations with Mordell-Weil group of
  rank $N-1$ or an $N$-section, which admit an axion if $N>3$.}

\begin{document}
\setlength{\parskip}{5pt}

\makeatletter
\renewcommand{\@fpheader}{%\old@fpheader
\hfill IFT-UAM/CSIC-14-130, MPP-2014-379}
\makeatother
%IFT-UAM/CSIC-14-130

\maketitle
\newpage

\section{Introduction}

The realization of inflationary models in string theory is a
long-standing challenge \cite{Baumann:2014nda}.  A large class of
models of inflation employ the dynamics of one or more scalar field
that slowly roll down a flat potential. While there has been progress
in understanding candidate scalar potentials that arise in string
theory \cite{Douglas:2006es,Blumenhagen:2006ci}, it remains
challenging to identify scalars with sufficiently flat potentials that
at the same time are the lightest scalar degrees of freedom during the
inflationary epoch.  In large field inflationary models this task
becomes even more demanding, since the flatness of the potential and
stability of other field space direction has to be controlled over
super-Planckian distances.  Large field inflationary models that
predict a large tensor-to-scalar ratio have recently gained much
attention due to the initial claim of the BICEP2 experiment to having
discovered primordial gravitation waves \cite{Ade:2014xna}. While this
result is still under debate \cite{Adam:2014bub}, it is in any case an
interesting conceptual task to realize large field inflation in string
theory.

Large field inflationary models can be constructed, for example, by
considering scalars that have an axion-like shift-symmetry in the
absence of a scalar potential.  This symmetry is then broken by a
scalar potential, which has to be controlled over a large field
range. Candidates for such axions are zero modes of the R-R and NS-NS
form fields of string theory. A scalar potential for these fields can
arise from brane or flux backgrounds or can be induced
non-perturbatively by brane-instantons or gaugino
condensates. Prominent scenarios implementing these steps are aligned
axion inflation models
\cite{Kim:2004rp,Kappl:2014lra,Long:2014dta,Gao:2014uha}, models of
N-flation
\cite{Liddle:1998jc,Dimopoulos:2005ac,Grimm:2007hs,Cicoli:2014sva,Bachlechner:2014hsa},
and models of axion monodromy inflation
\cite{Silverstein:2008sg,McAllister:2008hb,Kaloper:2008fb,Palti:2014kza,Marchesano:2014mla,Hebecker:2014eua,Blumenhagen:2014gta,Ibanez:2014kia,Arends:2014qca,McAllister:2014mpa,Franco:2014hsa,Ibanez:2014swa}. While
the necessary ingredients for these models are present in string
theory, the explicit realization of inflation in combination with
moduli stabilization is challenging. The situation improves if the
effective theory implementing the model preserves $\cN=1$
supersymmetry with a scalar sector described by a K\"ahler potential
and superpotential. For example, models of axion monodromy inflation
where suggested to also arise in supersymmetric theories from an
F-term breaking \cite{Marchesano:2014mla,Blumenhagen:2014gta}.  For
all of these scenarios, it should be stressed that the shift
symmetries of the fields for the kinetic terms are only present at
special points in moduli space, at which the extended objects coupling
to the form-fields are sufficiently heavy.

In this paper we study a rich class of axion inflationary models
arising from the axions being realized in the complex structure moduli
space of the internal manifold.  We consider mainly Type IIB
orientifold compactifications based on Calabi-Yau threefolds
\cite{Grimm:2005fa,Douglas:2006es,Blumenhagen:2006ci}, but also
comment on the generalization to F-theory on elliptically fibered
Calabi-Yau fourfolds \cite{Vafa:1996xn,Denef:2008wq}.  For such
Calabi-Yau geometries we suggest that one can systematically identify
special points in their complex structure moduli space at which
scalars exist that have approximately shift-symmetric kinetic
terms. In fact, these special points are often connected by dualities
to string theory setups in which the axions are R-R or NS-NS
form-field zero modes.  One example, which has recently been
investigated in this context
\cite{Hebecker:2014eua,Blumenhagen:2014nba,Hayashi:2014aua,Hebecker:2014kva}
is the `large complex structure point' in the moduli space of the
Calabi-Yau geometries. For Calabi-Yau threefold compactifications of
the Type II string theories one can use mirror symmetry to identify
the large complex structure point with the large volume point of a
dual geometry. In this situation the classical shift symmetry of the
NS-NS two-form is the dual source of the shift symmetry. With this
understanding one can also gain more control over the expected sources
that break the shift-symmetry. For example, at the large volume point
of Calabi-Yau threefolds one can easily see that the continuous
symmetry will be broken by world-sheet instantons.

In order to identify points in the complex structure moduli space at
which axions exist, we propose the following strategy: the complex
structure of a Calabi-Yau $n$-fold can be parameterized by the
integrals of the holomorphic $(n,0)$ form over a integral symplectic
basis of $H_n(X_n,\bZ)$. These integrals are known as the
\emph{periods} of $X_n$. At certain special points in complex
structure moduli space these periods can have orbifold or $\log$
singularities. One can then encircle these points and study the
behavior of the periods, which are typically not single-valued but
rather admit \textit{discrete monodromies} when encircling these
points. These discrete symmetries, however, do in general not suffice
to infer the presence of axions with \textit{approximate continuous
  shift-symmetries}.  We therefore have to further constrain our
considerations to special points in complex structure moduli space for
which the local monodromy transformation ensures that indeed the
K\"ahler potential admits a shift symmetry. We will present some
conditions on the local monodromy that allows for approximate shift
symmetries to exist.

A scalar potential for the complex structure moduli in Type IIB
Calabi-Yau orientifold and F-theory compactifications is induced upon
switching on background fluxes
\cite{Douglas:2006es,Blumenhagen:2006ci}. More precisely, one can show
that this potential arises from a superpotential depending on the
values of the periods \cite{Gukov:1999ya,Giddings:2001yu}. For
sufficiently generic fluxes the superpotential breaks the discrete
monodromy symmetry as well as the local continuous shift symmetry.
Evaluating this superpotential at different special points that admit
axions, we show that different types of shapes of scalar potentials
are induced. For example, while the potential is of polynomial-type
near the large complex structure point, it is of cosine-type near a
conifold point. Therefore a number of different inflation models arise
naturally from the effective theories that we get when we move close
to different special points in the moduli space.

To illustrate our proposal we will investigate several Calabi-Yau
threefold examples in detail and compute the explicit moduli
dependence of the periods at special points of the moduli space. A
first class of examples are one-parameter Calabi-Yau manifolds. These
have been investigated in the study of mirror symmetry before in
\cite{Cox:2000vi,Hori:2003ic,Aspinwall:2004jr,Berglund:1993ax,Hosono:1993qy,Hosono:1994ax,Greene:2000ci,Mayr:2000as}. A
second class of examples are Calabi-Yau threefolds that are mirror
dual to elliptic fibrations. We show that these admit special points
admitting axions if the elliptic fibration has either Mordell-Weil
group of rank $N-1$ or an $N$-section with $N>3$. Near a certain
`small complex structure point' we identify the axion in the resulting
models of \textit{Mordell-Weil inflation}.

Before starting our investigations let us stress that our aim is not
to construct completely realistic inflation models, since this would
require a detailed study of moduli stabilization. We rather hope to
argue that the complex structure moduli space of Calabi-Yau manifolds
is rich enough to offer all the necessary building blocks to engineer
a large class of phenomenologically appealing axion inflation
models. It also allows to compute corrections and study stability of
the vacua.

The paper is organized as follows. In section \ref{general_story} we
first briefly introduce Type IIB orientifold setups and their F-theory
generalizations.  We discuss the form of the K\"ahler potential and
flux superpotential with particular focus on the complex structure
moduli dependence.  This will allow us to describe the basic idea to
identify axions in this moduli space. We will also give already a
brief summary of the results with detailed computations carried out in
sections \ref{special_oneparameter} and \ref{special_MW}.  In section
\ref{special_oneparameter} we study periods of one-parameter
Calabi-Yau threefolds in detail and provide the computation of the
K\"ahler potential and flux superpotential at various special points
in moduli space. In the final section \ref{special_MW} we turn to the
study of the complex structure moduli space for mirror threefolds of
elliptic fibrations.

\section{Inflation at special points in complex structure moduli space} \label{general_story}

In this section we introduce our setup and describe the key steps to
identify axions in complex structure moduli space. The Type IIB
orientifold compactifications on Calabi-Yau threefolds will be
discussed in subsection \ref{ori-setup}.  We introduce the general
form of the K\"ahler potential and flux superpotential with focus on
the complex structure moduli. A discussion of the monodromy group
generated due to special points in moduli space allows us to state the
geometric requires for the existence of axions. Some of our main
results on the resulting form of the K\"ahler potentials and flux
superpotentials at certain special points in moduli space are
described in subsection \ref{threefoldresults}. Finally, we comment on
the generalization to F-theory compactifications on fourfolds in
subsection \ref{F-theorygen}.

\subsection{Axions at special points in orientifold set-ups} \label{ori-setup}

Most of our discussion will take place in the context of Type IIB Calabi-Yau 
orientifold compactifications with $O7^-$ planes. In order to cancel tadpoles these 
set-ups will also include space-time filling D7-branes and fluxes. 
The resulting system can preserve (possibly spontaneously
broken) $\cN=1$ supersymmetry, we will be focusing on situations in
which this is the case. Such compactifications allow many
appealing features from the point of view of string model building
\cite{Douglas:2006es,Blumenhagen:2006ci}. They also arise as the weak string coupling limit 
of F-theory compactifications as we will briefly discuss in section \ref{F-theorygen}.
 
Before recalling aspects of the $\cN=1$ effective theory 
relevant to this work, let us quickly summarize some facts about 
the construction of orientifold models following \cite{Giddings:2001yu,Grimm:2004uq}.
The starting point is a Calabi-Yau threefold $X_3$ admitting a holomorphic $\bZ_2$ involution
$\sigma\colon X_3\to X_3$. We demand that $\sigma$ acts on the holomorphic $(3,0)$-form
$\Omega$ on $X_3$ as 
\begin{equation} \label{Omegatrans}
   \sigma^*\Omega = -\Omega\ .
\end{equation} 
The physical
model is constructed by quotienting Type IIB string theory on $X_3$ by
$(-1)^{F_L}\sigma\, \Omega_p$, with $\Omega_p$ the orientation reversal action
on the worldsheet, and $F_L$ the left-moving fermion number. 
The fixed loci of such an involution are divisors of
$X_3$, which we identify with $O7$ planes, and possibly points in
$X_3$, which get identified with $O3$ planes.

The general four-dimensional $\cN=1$ effective theory for the bulk 
moduli of such 
orientifold set-ups has been worked out in \cite{Grimm:2004uq}. 
Supersymmetry implies that the dynamics of the fields 
can be encoded by a K\"ahler potential $K$ and a superpotential $W$. 
In the following we will focus on the moduli sector of such theories. 
More precisely, we will denote by $\tau = C_0 + i e^{-\phi}$ 
the dilaton-axion field, and by $z^k$ the $h^{2,1}_-(X_3)$ 
complex structure moduli compatible with \eqref{Omegatrans}. 
The $(3,0)$-form $\Omega$ depends on the complex structure 
moduli $z^k$. At classical order, the K\"ahler potential for $\tau, z^k$ 
takes the form
\begin{equation}
  \label{eq:K}
  K  = -\log\big[ i(\tau - \ov \tau)\big] - \log\left[i \int_{X_3} \Omega\wedge \bar \Omega
  \right]+\ldots
\end{equation}
where the dots indicate terms depending on other moduli or matter fields of the 
set-up. A superpotential is induced by R-R and NS-NS background fluxes 
$F_3,H_3$ \cite{Gukov:1999ya,Giddings:2001yu}.  Defining $G_3=F_3-\tau H_3$ it takes the form
\begin{equation}
  \label{eq:W}
  W = \int_{X_3} G_3\wedge \Omega + \ldots
\end{equation}
where the dots denote non-perturbative corrections that we will not
discuss in detail in this work.  Let us stress that in principle one
has to control all other terms in \eqref{eq:K} and $\eqref{eq:W}$ in
building an inflationary model with complete moduli
stabilization. This is expected to be challenging, since the
$z^k,\tau$ are known to mix with other fields in the suppressed terms
in \eqref{eq:K} and \eqref{eq:W}.  Recent more complete studies of
moduli stabilization in related settings can be found in
\cite{Blumenhagen:2014nba,Hayashi:2014aua,Hebecker:2014kva}. To convey
our message about the existence of axions and the varying shapes of
the scalar potential we will assume that we can study the dynamics of
the $z^k$ using the displayed terms in \eqref{eq:K} and
$\eqref{eq:W}$.  Furthermore, we will consider the case
$h^{2,1}_+(Y_4) =0$, such that the orientifold projection does not
complicate our study additionally. This condition can be weakened, the
generic situation can be naturally studied in the F-theory setting
discussed in section~\ref{F-theorygen}.

We study in this work inflationary models for which the inflaton is 
among the complex structure moduli $z^k$. It will be therefore crucial 
to examine the dependence of $\Omega$ on $z^k$. 
In order to do that it is convenient to choose  
a basis $\cA_i$ for $H_3(X_3,\bZ)$ and define the 
periods 
\begin{equation}
   \Pi^i = \int_{\cA_i} \Omega\ ,
\end{equation}
Clearly, the $\Pi^i$ are depending on the fields $z^k$ through $\Omega$. 
Furthermore, the periods transform under $Sp(2(h^{2,1}+1))$, 
which gives the freedom to choose a symplectic basis $\cA_i$ with 
\begin{equation}
   \cA_i \cap \cA_j = \eta_{ij}\ , \qquad  \eta_{ij} = \left(\begin{array}{cc}0 & \mathbf{1}\\  -\mathbf{1}& 0\end{array}\right)\ ,
\end{equation}
where $\mathbf{1}$ is the $(h^{2,1}+1) \times (h^{2,1}+1)$ unit matrix. 
The action of the group $Sp(2(h^{2,1}+1))$ on the periods $\Pi^i$ is 
via 
\begin{equation} \label{Pirot}
   \Pi'^i = T_{\ j}^i \, \Pi^j\ . 
\end{equation}
keeping the K\"ahler potential invariant.
We stress though that \eqref{Pirot} is in general not a symmetry of the system. There is, however, a concept of 
symmetry group $G_{\rm mon} \subset Sp(2(h^{2,1}+1))$, known as the monodromy group, which we will introduce below.
To evaluate its action on the physical quantities we 
note that $K$ and $W$ given in \eqref{eq:K}, \eqref{eq:W} can be written as
\begin{align}
  K  &= -\log\big[ i (\tau - \ov \tau) \big]- \log \big[i \Pi^i \eta_{ij} \bar \Pi^j   \big]+\ldots  \label{Kinperiods} \\[.1cm]
  W & = \cN_i \Pi^i + \ldots \ , \qquad  \cN_i = N_i - \tau M_i\ ,  \label{Winperiods}
\end{align}
where $N^i = \int_{\cA_i} F_3$ and $M^i = \int_{\cA_i} H_3$ are the flux quanta. 

A special role will be played by the discrete group $G_{\rm mon}$. 
This group is an actual symmetry of the K\"ahler potential terms displayed in \eqref{Kinperiods}
independent of the considered point in complex structure moduli space.
To introduce this group we first note that the moduli space 
of complex structure generally admits special points that are given by the 
loci at which some of the periods $\Pi^i$ become singular.
If one marks these points in the moduli space one can now encircle them and investigate how the 
periods transform. Let us consider one such a point $z_{\rm s}$  and denote 
the matrix providing a symmetry transformation of the periods by $T_{\ i}^j [z_{\rm s}]$, i.e.
\begin{equation}
    T_{\ i}^j[z_{\rm s}] \,\Pi^i(z) =  \Pi^j(z)\ . 
\end{equation}
Collecting all such $ T_{\ i}^j [ z_{\rm s}]$ for all special points, one can show that they form a 
group $G_{\rm mon}\subset Sp(2(h^{2,1}+1))$. This monodromy group and special 
points will play the key role in this work. 

\begin{figure}[h!]
\centering
\vspace*{-.5cm}
\begin{picture}(100,100)
\put(-90,0){\includegraphics[width=0.45\textwidth]{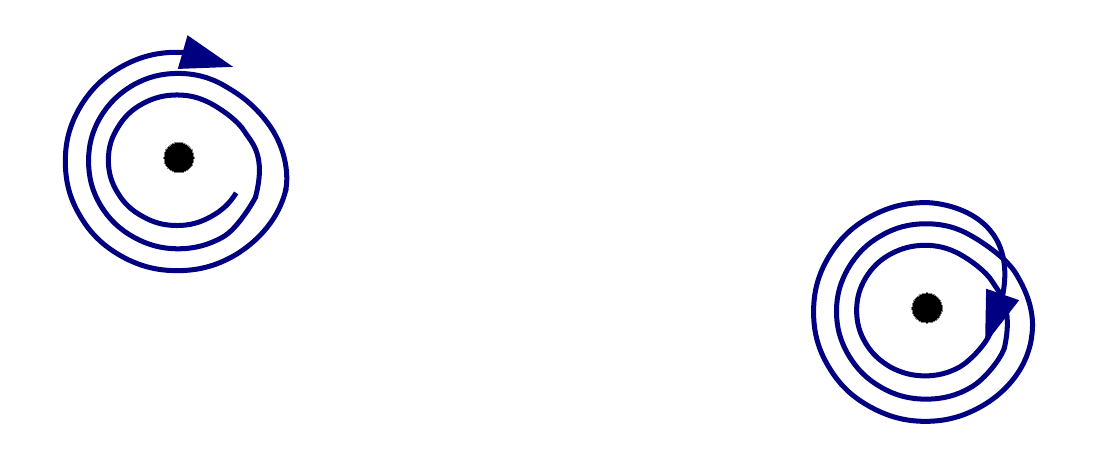}}
\put(90,50){$T[z_2]^n = T[z_2]$}
\put(-40,30){$z_1$}
\put(95,5){$z_2$}
\end{picture}
\begin{minipage}{12cm}
\caption{Schematic depiction of singular points in complex structure 
moduli space with infinite-order monodromy $z_1$ and finite-order monodromy $z_2$. An axions can be identified 
very close to the point $z_1$.} \label{fig:finiteinfinite}
\end{minipage}
\end{figure}
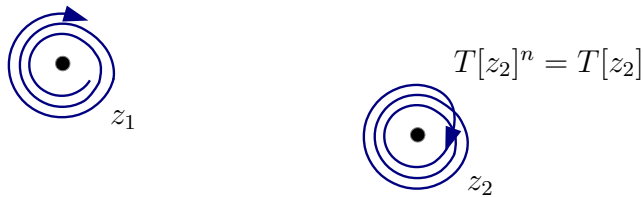

To be more precise, our interest in this paper is on identifying candidates for
\emph{large-field} inflationary models that arise close
to the special points $z_{\rm s}$ in complex structure moduli space.
Our strategy is given by 
\begin{itemize}
  \item  We aim to identify points in the moduli space near which the
    K\"ahler potential $K$ given in \eqref{eq:K} 
    has an approximate continuous shift symmetry, coming
    from going around the marked point in moduli space. So we have a
    natural candidate for an \textit{axion} $\theta$, which we will able to identify in specific examples. 
    We claim that this arises, in particular, at special points $z_{\rm s}$ for which the 
    monodromy group element $T_{\ i}^j [z_{s}] \in G_{\rm mon}$ acting on the periods is
    \emph{of infinite order}, i.e.~there exists no $n$ such that $T[{z_s}]^n = T[{z_s}]$.\footnote{Let us stress that only in the case of infinite order monodromies 
we are able to directly identify an axion. Nevertheless it would be interesting 
to investigate finite order situations with finite order monodromy from a
phenomenological point of view. } The leading 
    order behavior of $K$ can then be computed using techniques for period 
    computations \cite{Cox:2000vi,Hori:2003ic,Aspinwall:2004jr,Berglund:1993ax,Hosono:1993qy,Hosono:1994ax,Greene:2000ci,Mayr:2000as}
    or the recent direct approach using localization \cite{Benini:2012ui,Doroud:2012xw,Jockers:2012dk,Halverson:2013qca,Hori:2013ika,Kim:2013ola}. 

  \item In a second step we then consider the flux induced
    superpotential $W$ given in \eqref{eq:W}. If the vectors $M^i,N^i$
    of chosen fluxes are not invariant under $T_{\ i}^j [z_{s}]$ the
    superpotential breaks the approximate shift symmetry of the
    K\"ahler potential spontaneously. In other words, there is a flux
    induced scalar potential for the axion $\theta$. The fact, that we
    restricted our considerations to monodromies of infinite order
    implies that in principle one can keep increasing the energy of
    the inflaton $\theta$ indefinitely. The two situations of finite
    and infinite order are depicted in
    Figure~\ref{fig:finiteinfinite}.
\end{itemize}

The main aim of this paper is to show that there are indeed rather
simple situations in which such special inflationary points in moduli
space arise. We will discuss two main families of examples in the 
later sections: 
(1) one-parameter Calabi-Yau models (i.e. Calabi-Yau threefolds with a
single complex structure deformation), and (2) Calabi-Yau threefolds 
with the mirror geometry being torus fibered, in which the problem, 
taking a limit in the
parameters describing the base, reduces effectively again to a
one-parameter model. We will see that in both cases there are
interesting special points, beyond the large complex structure limit
previously studied in detail \cite{Hebecker:2014eua,Hayashi:2014aua,Blumenhagen:2014nba,Hebecker:2014kva}. 

\subsection{Summary of results for Calabi-Yau threefolds} \label{threefoldresults}

While presenting the details of the individual models will require 
to introduce more mathematics, as done in sections \ref{special_oneparameter} and \ref{special_MW}, 
it is instructive to already have a first look at the results. In particular, 
one indeed finds that very different inflationary potentials arise at different 
points in moduli space. Examples are the `large complex structure point', the `small complex structure point'
and the `conifold point'. We will 
discuss the various results on these points in the following

\begin{itemize}
\item \textit{Large complex structure point:} The analysis of the
  large complex structure point is related by mirror symmetry to the
  large volume behavior of a mirror-dual Calabi-Yau threefold $Y_3$
  to our original manifold $X_3$.  We will stepwise introduce the
  systematics for computing the periods expanded around the large
  complex structure point in section \ref{special_oneparameter} and
  show that the monodromy is of infinite order.  Restricting to
  one-parameter models, one indeed finds an axion that upon choosing
  appropriate coordinates can be identified with the real part of a
  complex scalar $t$. The large complex structure point is located at
  $t = i \infty$.  The K\"ahler potential $K$ and flux superpotential
  $W$ expanded around this point in moduli space take the form
  \begin{align} \label{KWlarge}
  K_{\rm lcs}&= - \log\big[ i (\tau-\bar \tau) \big]- \log \left[i\left(-\frac{1}{6} \cK (t-\bar{t})^3+ 2 \hat c\right)\right] \ ,  \\[.1cm]
  W_{\rm lcs} &= \frac{1}{6} \cN_4 \Kint t^3-\frac{1}{2}\cN_3 \Kint
  t^2+\left(\cN_4\hat b+\frac{1}{2}\cN_3
    \Kint+\cN_2\right)t+\left(\cN_1-\cN_4\hat{c}+\cN_3\hat{b}\right) \
  , \nn \end{align} where $\cN_i$ was defined in \eqref{Winperiods} and $\hat
  c = \frac{\zeta(3)\chi}{(2\pi i)^3}$.  The topological numbers
  $\cK$, $\hat{b}$, and $\chi$ of the mirror geometry $Y_3$ to $X_3$
  will be introduced in \eqref{eq:bhat}.  Let us note that
  \eqref{KWlarge} yields the polynomial-type potentials that have been
  discussed in more detail in
  \cite{Marchesano:2014mla,Hebecker:2014eua,
    Hayashi:2014aua,Blumenhagen:2014nba,Hebecker:2014kva}.

\item \textit{Small complex structure point:} A second special point
  in one-parameter models is known as the small complex structure
  point. Upon choosing an appropriate complex coordinate $u$, with the
  small complex structure point located at $|u|=\infty$, we aim to
  identify the axion with the phase of the complex coordinate $u$.  It
  should be stressed, however, that at this special point the phase
  has not necessarily an approximate shift-symmetry for each
  geometry. Concretely, we consider a geometry $X_3$ with mirror
  geometry being the complete intersection $Y_3=\bP^n[d_1\ldots d_k]$,
  meaning the complete intersection of the hypersurfaces of degree
  $d_1,\ldots,\d_k$ in $\bP^n$. This allows to define the set
  $\{\alpha_i\}=\{1/d_1, \ldots, (d_1-1)/d_1,
  1/d_2,\ldots,(d_k-1)/d_k\}$.  One now checks that only if at least
  two $\alpha_i$ coincide, one actually finds a monodromy matrix of
  infinite order and therefore an axion.  In this case, the leading
  K\"ahler potential and superpotential expanded around such points in
  moduli space take the schematic form
\begin{align} \label{KWsmcs}
K_{\rm scs} &= - \log\big[ i (\tau-\bar \tau) \big] -  \log\left[a|u|^{-2\alpha_\kappa} \log|u|+\dots\right] \ ,  \\
W_{\rm scs}&=\sum_i N^{\text{eff}}_i u^{-\alpha_i}(\log(u)+\dots)+\sum_j N'^{\text{eff}}_ju^{-\alpha_j}+\dots \ .  \nn 
\end{align}
This expansion displays only the leading terms. In particular,
$\alpha_\kappa$ is the smallest repeated constant in the set
$\{\alpha_i\}$. The first sum in $W_{\rm scs}$ is running over the
repeated $\alpha_i$ and the second sum over the non-repeated ones, as
we will explain in more detail in section
\ref{special_oneparameter}. The parameter $a$ is a complex function of
the $\alpha_i$ and the topological numbers of $Y_3$, while the
$N^{\text{eff}}_i,\, N'^{\text{eff}}_j$ are linear combinations of the
fluxes $\cN_1,\ldots,\cN_4$ with coefficients dependent on the
$\alpha_i$ and topological numbers $Y_3$. (We will fully determine
these quantities in our examples in section~\ref{special_oneparameter}.)

\item \textit{Conifold point:} A third special point of interest is the so-called conifold point. 
In a one-parameter model one can choose a coordinate $t_c$ such that it is located 
at $|t_c| = 0$, with the axion being the phase $\theta$ of $t_c$. 
Expanding the K\"ahler potential and superpotential around this point one finds 
\begin{align} \label{KWcon}
K_{\rm con} &=  - \log\big[ i (\tau-\bar \tau) \big] - \log\left[\frac{-1}{\pi} \Kint |t_c|^2 \log|t_c|+\dots\right] \ ,  \\[.1cm]
W_{\rm con}&=-\cN_4 \Kint t_c \log(t_c)+\frac{1}{2} \cN_3 \Kint t_c^2+\big(\cN_2-\frac{1}{2}\cN_4 \cK\big)t_c+\cN_1\ .  \nn 
\end{align}
It is interesting to point out that the scalar potential derived from
$K_{\rm con},\ W_{\rm con}$ admits $\cos (\theta)$ as a leading term,
i.e.~a periodic potential used in models of natural inflation
\cite{Freese:1990rb,Adams:1992bn}. This is not unexpected, since, at
least in geometries with more than one complex structure modulus, one
can sometimes perform a geometric transition replacing the fluxes with
a stack of five-branes
\cite{Vafa:2000wi,Cachazo:2001jy,Dijkgraaf:2002dh,Heckman:2007ub,Aganagic:2007py}. These
branes induce a non-perturbative superpotential due to a gaugino
condensate resulting in a $\cos(\theta)$ term in the scalar potential.

\end{itemize}
While a detailed phenomenological analysis of the various occurring
potentials is beyond the scope of this paper, it is intriguing to
realize that already the simple examples above yield different
inflationary potentials.

Let us close this section by also commenting on some 
generic special points that arises for Calabi-Yau threefolds $X_3$ for 
which the mirror geometry $Y_3$ is torus-fibered. 
Such torus-fibered $Y_3$ admit generically a new type 
of special locus in their K\"ahler moduli space at which the corresponding
geometries $Y_3$ admit a torus fiber with volume shrunken to zero. In the mirror $X_3$ this corresponds
to a special locus in the complex structure moduli space. While 
the general analysis of the K\"ahler potential and superpotential near this locus
is rather involved, one can further consider the limit in which the base 
of $Y_3$ is large, i.e.~$Y_3$, $X_3$ behave in the base direction as in 
the discussion of the `large complex structure point' above.   
This mixed limit yields a special point in the complex structure 
moduli space of $X_3$ that we term the `F-point'.\footnote{We 
note that this limit coincides on the K\"ahler moduli side with the 
F-theory limit of an M-theory setup \cite{Denef:2008wq,Grimm:2010ks}.} 
\begin{itemize}
\item \textit{F-point:}  Studying the monodromy around the F-point, we will see in section \ref{special_MW}  
that the existence of axions at this point depends on the structure of sections of $Y_3$. 
More precisely, if $Y_3$ admits an elliptic fibration with $N$ sections, 
i.e.~the Mordell-Weil group has rank $N-1$, then an axion exists for $N>3$. 
Alternatively, if $Y_3$ has an $N$-section the same bound $N>3$ applies
for having an axion. The K\"ahler potential and superpotential for the mirrors of elliptic fibrations 
can be evaluated at various points in moduli space similar to the approach 
of section \ref{special_oneparameter}.\footnote{A recent discussion on mirror symmetry
for elliptic fibrations can be found in \cite{Alim:2012ss,Klemm:2012sx}.} Remarkably,
when $N>3$ the periods near the F-point admit a new symmetry that 
allows to map their structure to the large complex structure point for 
both base and fiber. Therefore, one expects that the leading terms 
of $K,\, W$ will be of the type \eqref{KWlarge}, but generalized to include more 
than one modulus. One furthermore finds that the coefficients of the various terms depend on the 
topological data of the torus fibered $Y_3$ in a distinguished way and will 
single out the axion associated to the torus fiber.
\end{itemize}

We will call the models arising near the F-point, \textit{Mordell-Weil inflation}.
While the physical setup is somewhat different to
ours, the mathematical techniques recently developed for the study of
F-theory compactifications with extra $U(1)$ 
symmetries \cite{Grimm:2010ez,Morrison:2012ei,Braun:2013yti,Cvetic:2013nia,Borchmann:2013jwa,Grimm:2013oga,Braun:2013nqa,Borchmann:2013hta,Cvetic:2013qsa,Braun:2014oya,Morrison:2014era,Kuntzler:2014ila,Klevers:2014bqa,Braun:2014qka,Lawrie:2014uya,Anderson:2014yva,Garcia-Etxebarria:2014qua,Mayrhofer:2014haa,Mayrhofer:2014laa} will be
directly applicable to determine the topological data required for the 
construction of inflationary models suggested here. It is an exciting task 
to carry out this analysis further and check if the constraints on moduli 
stabilization recently found in \cite{Hayashi:2014aua,Blumenhagen:2014nba,Hebecker:2014kva} can be met in these setups.

Let us close this section by noting that the functional forms for $K,W$ encountered in 
\eqref{KWlarge}, \eqref{KWsmcs}, and \eqref{KWcon} are not the only 
possibilities of situations that might occur, but rather are the results 
of our explicit analysis in sections \ref{special_oneparameter} and \ref{special_MW}.
It is an interesting task to extend the list of special points and study other candidate
K\"ahler potentials and superpotentials. This becomes particularly interesting 
when extending the analysis to Calabi-Yau fourfolds use in the F-theory generalization
discussed next.
 
\subsection{Generalization to F-theory} \label{F-theorygen}

In this section we discuss the generalization of the orientifold construction of 
section~\ref{ori-setup} to F-theory. The F-theory set-up 
of interest are compactifications on Calabi-Yau fourfolds $X_4$ that 
are elliptically fibered. We will argue, on the one hand, 
that the strategy to identify axion-like fields is very similar to the 
orientifold approach and amount to a detailed study of the 
complex structure moduli space of $X_4$. The explicit computations are, however, mathematically 
and technically more involved \cite{Mayr:1996sh,Klemm:1996ts,Alim:2009bx,Grimm:2009ef,Bizet:2014uua}. On the other hand, the F-theory approach is a significant 
generalization. In order to appreciate this, we stress that the complex structure 
moduli of $X_4$ capture not only the degrees of freedom of $\tau$ and the 
complex structure moduli of $Y_3$, but also the seven-brane moduli. Therefore, 
it should allow to include the inflationary models considered in \cite{Hebecker:2011hk,Hebecker:2012aw,Hebecker:2014eua}. 

To begin with we recall some facts about the complex structure dependent 
K\"ahler potential and flux superpotential of an F-theory compactification 
on $X_4$. Let us denote by $\Omega_4$ the $(4,0)$-form on $X_4$, 
which is known to vary holomorphically over the complex structure moduli 
space with local coordinates $z^I$.  Classically the K\"ahler potential 
for $z^I$ takes the form
\begin{equation}
  \label{eq:K4}
  K^{\rm F}  = - \log\left[i \int_{X_4} \Omega_4 \wedge \bar \Omega_4
  \right]+\ldots
\end{equation}
where the dots again indicate terms depending on other moduli or matter fields of the 
set-up. The superpotential is now induced by a real 4-form flux $G_4$ and takes the 
form
\begin{equation}
  \label{eq:W4}
  W^{\rm F} = \int_{X_4} G_4\wedge \Omega_4 + \ldots
\end{equation}
where the dots indicate possible non-perturbative corrections.  The
K\"ahler potential \eqref{eq:K4} and superpotential can be obtained by
taking the M-theory to F-theory limit
\cite{Denef:2008wq,Grimm:2010ks}. In general, both are complicated
functions of the moduli $z^I$, which can, however, be evaluated for
certain given smooth Calabi-Yau fourfolds. Both \eqref{eq:K4} and
\eqref{eq:W4} are true generalizations of the weak coupling
counterparts \eqref{eq:K}, \eqref{eq:W} not only because they depend
also on the seven-brane moduli, but also because they capture
information about the $\tau$ expansion, and hence the $g_s$
corrections, beyond the leading terms present at large $\I(\tau)$.  It
should be stressed, however, that particularly the K\"ahler potential
is likely to admit several perturbative and non-perturbative
corrections and it remains a challenging task to examine these in
detail.

As in the weak coupling setups of section \ref{ori-setup} one can introduce 
the periods
\begin{equation}
   \Pi^\cI(z) = \int_{\Gamma_\cI} \Omega_4\ , 
\end{equation} 
where $\Gamma_\cI$ is a basis of four-cycles of $H_{4}(X_4,\mathbb{Z})$,
with intersection product 
\begin{equation} 
   \eta_{\cI \mathcal{J}} = \Gamma_\cI \cap \Gamma_\mathcal{J}\ .
\end{equation}
In terms of these quanteties one expresses \eqref{eq:K4} and \eqref{eq:W4}
as 
\begin{equation}
  K  = - \log \big[i \Pi^\cI \eta_{\cI \mathcal{J}} \bar \Pi^{\mathcal{J}}   \big]+\ldots  \ , \qquad
  W  =N_\cI \Pi^\cI + \ldots \ ,   
\end{equation}
where $N^\cI = \int_{\Gamma_\cI} G_4$ are the flux quanta. Despite the 
similarities with \eqref{Winperiods}, the proper treatment 
of the periods on Calabi-Yau fourfolds is significantly 
more involved \cite{Mayr:1996sh,Klemm:1996ts,Alim:2009bx,Grimm:2009ef,Bizet:2014uua}.
This can be traced back to the fact that there is no underlying special geometry, as it is present
for Calabi-Yau threefolds, that dictades already key features of the couplings.

It should be stressed that the Calabi-Yau fourfolds $X_4$ used in order to describe 
an F-theory compactification have to admit an elliptic fibration, or rather a two-torus 
fibration. Furthermore, it is often the case that also the mirror dual geometry $Y_4$ 
to $X_4$ admits an elliptic fibration. Let us assume that we have a pair $X_4$, $Y_4$
of mirror manifolds, which are both elliptically fibered.
For both geometries one can then introduce the Mordell-Weil groups $\text{MW}(X_4)$ 
and $\text{MW}(Y_4)$. On the one hand, the 
rank of $\text{MW}(X_4)$ is giving the number of massless $U(1)$ gauge fields 
in the effective theory, and therefore a study of $\text{MW}(X_4)$ is of key importance
in F-theory model building \cite{Grimm:2010ez,Morrison:2012ei,Braun:2013yti,Cvetic:2013nia,Borchmann:2013jwa,Grimm:2013oga,Braun:2013nqa,Borchmann:2013hta,Cvetic:2013qsa,Braun:2014oya,Morrison:2014era,Kuntzler:2014ila,Klevers:2014bqa,Braun:2014qka,Lawrie:2014uya,Anderson:2014yva,Garcia-Etxebarria:2014qua,Mayrhofer:2014haa,Mayrhofer:2014laa}.
On the other hand, the rank of $\text{MW}(Y_4)$ has no known physical 
meaning in the effective theory. However, as for the Calabi-Yau threefolds, one can 
check that if the rank $N-1$ of $\text{MW}(Y_4)$ is sufficiently large, then the complex structure 
moduli space will admit an axion at the special F-point introduced in subsection \ref{threefoldresults}. 
Indeed, for $N>3$ the monodromy around this point turns out to be of infinite order and one encounters 
models of \textit{Mordell-Weil inflation}. It would be very exciting to 
examine thoroughly if the presence of such axion points in complex structure moduli 
space is generic in F-theory geometries.

\section{Special points and axions in one-parameter threefolds}\label{special_oneparameter}

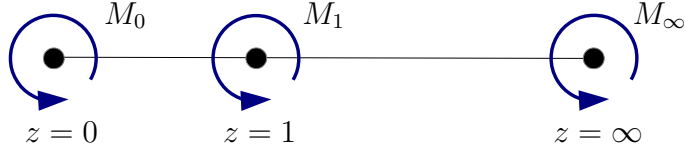
\begin{figure}[h!]
\centering
\vspace*{-.5cm}
\begin{picture}(100,100)
\put(-90,0){\includegraphics[width=0.6\textwidth]{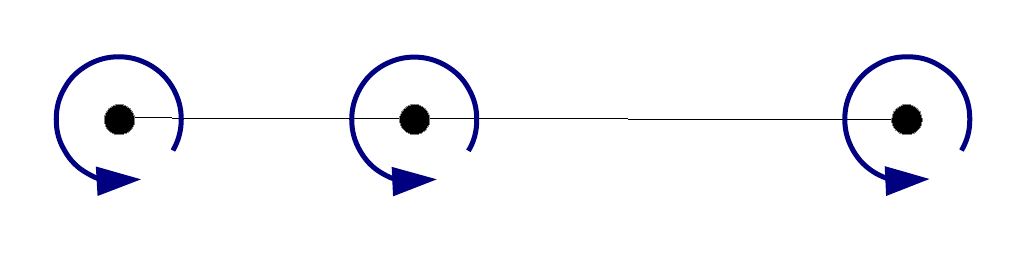}}
\put(5,5){$z=1$}
\put(35,50){$M_1$}
\put(-70,5){$z=0$}
\put(-40,50){$M_0$}
\put(130,5){$z=\infty$}
\put(160,50){$M_\infty$}
\end{picture}
\begin{minipage}{12cm}
  \caption{Schematic depiction of the one-parameter moduli space of
    Calabi-Yau threefolds with three special points.  $M_0$, $M_1$,
    and $M_\infty$ denote the monodromy matrices around these
    points.} \label{fig:OneParameter}
\end{minipage}
\end{figure}

In this section we systematically study the special points in the
complex structure moduli space of Calabi-Yau threefolds. To be
explicit in our examination we restrict to Calabi-Yau manifolds $X_3$
with one complex structure modulus $z$, i.e.~we consider geometries
with $h^{2,1}(X_3) = 1$.  As described in section~\ref{general_story}
the $\cN=1$ K\"ahler potential and superpotential are determined by
the periods of $\Omega$.  They can undergo a monodromy transformations
about special points of the geometry.  In one parameter models the
candidates to be special points are the large complex structure point,
the small complex structure point, and the conifold point. We
parametrize the moduli space such that these points are located at
$z=0$, $z=\infty$, and $z=1$, respectively, see
figure~\ref{fig:OneParameter}.  The aim of this section is to
formulate a systematic approach to identifying the axion with a shift
symmetry from the monodromy transformations present about these
different special points in the complex structure moduli space.  We
will see that this also allows to systematically derive the
inflationary potential, which turns out to be fully determined by some
simple topological invariants of the manifold under study.

For the sake of concreteness we will restrict ourselves in this
section to studying one-modulus Calabi-Yau threefolds $X_3$ constructed
as mirrors of complete intersections on an ambient projective
space.\footnote{We will use well known ideas in the context of mirror
  symmetry, so our discussion will be somewhat concise at points. For
  nice reviews of the background material we refer the reader to
  \cite{Cox:2000vi,Hori:2003ic,Aspinwall:2004jr}.} This mirror dual to
$X_3$ will be denoted by $Y_3$. Our ideas generalize easily to more
involved configurations, using the techniques described in
\cite{Hosono:1993qy,Hosono:1994ax}. It is a well known fact that
periods of the holomorphic three-form $\Omega$ in such spaces satisfy
a differential equation known as the Picard-Fuchs equation, which we
now describe. Consider a complete intersection $Y_3=\bP^n[d_1\ldots
d_k]$, where the notation means the complete intersection of $k$
generic homogeneous polynomials of degrees $d_1,\ldots,d_k$ in
$\bP^n$. In order for $Y_3$ to be Calabi-Yau we need that $\sum_{i=1}^k
d_i=n+1$, and in order for it to be a threefold we need $n-k=3$. It
was proven in \cite{Hosono:1994ax} that the periods $\Pi$ on $X_3$
satisfy the Picard-Fuchs equation
\begin{equation}
  \wt{\cL}_d\Pi(z) \equiv \left(\theta^{n+1} -
    \left\{\prod_{i=1}^k(d_i\theta)(d_i\theta - 1)\cdots(d_i\theta -
      d_i + 1)\right\}z \right)\Pi(z) = 0
\end{equation}
with $z$ a complex structure parameter for $X_3$ (with $z=0$
corresponding to large volume in $Y$), and $\theta=z\partial/\partial
z$. Notice that we can always factorize $\wt{\cL}_d=\theta^k\cL_d$, so
we can find periods by solving the simpler equation
\begin{equation}
  \cL_d\Pi(z) = \left(\theta^4 - z \prod_{i=1}^k d_i(d_i\theta +
    1)\cdots(d_i\theta + d_i-1)\right)\Pi(z) = 0\, .
\end{equation}
We have commuted $z$ to the left for later convenience. Notice that
both terms in $\cL_d$ are fourth order in $\theta$ for Calabi-Yau
threefolds. We can make this more manifest by writing
\begin{equation}
\label{alphai}
  \cL_\alpha\Pi(z) \equiv \left(\theta^4 - \kappa z
    \prod_{i=1}^4(\theta + \alpha_i )\right) \Pi(z) = 0
\end{equation}
with $\kappa=\prod_i d_i^{d_i}$, and $\{\alpha_i\}_{i=1,\ldots, k}=\{1/d_1, \ldots,
(d_1-1)/d_1, 1/d_2,\ldots,(d_k-1)/d_k\}$. It is then convenient to
introduce $u=\kappa z$ and finally write
\begin{equation}
  \label{eq:PF}
  \cL_\alpha\Pi(u) = \left(\theta^4 - u \prod_{i=1}^4(\theta +
    \alpha_i )\right) \Pi(u) = 0\, .
\end{equation}
In what follows we will use the $u$ and $z$ variables interchangeably,
depending on what is most convenient.

\subsection{Analysis of monodromies}

\label{sec:monodromies}

Before going into technical details, let us comment on the monodromy
transformations found close to the different special points in the
complex structure moduli space of Calabi-Yau threefolds. Near the
large complex structure point ($z=0$) and the conifold point ($z=1$)
the monodromy transformations are unipotent matrices. By definition
this means that $(T-I)^{m-1}\neq 0$ and $(T-I)^{m}= 0$ for some $m$
called the index, and $T$ being the monodromy matrix. These matrices
are automatically of infinite order, i.e. there exists no $n$ such that
$T^n=I$. As we already mentioned in section \ref{general_story} this
fact implies the presence of a natural axion with an approximate
continuous shift symmetry in the four-dimensional effective
theory. More in general, around the special points we study in this
paper one has a monodromy matrix satisfying $(T^p - I)^m=0$. We want
to stress here that the qualitative form of the scalar potential in
the effective theory seems to be, to a large extent, determined by the
properties of the monodromy matrix alone, in particular by the value
of the indexes $m$ and $p$, although we have not worked out the
general dictionary. It would be interesting to go further in this
direction and look for a direct and systematic way of extracting from
the monodromy matrix the information about the low energy physics of
the axion, without following the somewhat painful route of computing
the periods that we take in this paper.

In particular, the solutions of \eqref{eq:PF} expanded near the large
complex structure point have unipotent matrix of index 4,
i.e. $(T[0]-I)^{4}= 0$. This implies (as we will see in more detail in
section \ref{sec:large}) that a natural basis of periods around this
point is given at leading order by $(1,\log(z),\log^2(z),\log^3(z))$,
giving rise to the four-dimensional effective theory already discussed
in section \ref{threefoldresults} which yields a polynomial-type
potential for the axion. In the examples that we study the monodromy
matrix near the conifold point is instead unipotent of index 2,
i.e. $(T[1]-I)^{2}= 0$, and the natural basis of periods is given by
$(1,t_c,t_c^2,t_c\log(t_c))$ with $t_c$ some appropriate local
coordinate. The scalar potential in this case turns out to acquire a
cosine-type form.

The small complex structure point is in this sense much more richer
than its partners, since the properties of the monodromy matrix depend
on the structure of the $\alpha_i$. Let us now advance the main result
that we will find: the existence of special points of infinite
monodromy around $z=\infty$ will depend on the structure of the
$\alpha_i$: if all the $\alpha_i$ are different the monodromy around
$z=\infty$ will be of finite order, and thus there will be no natural
candidate for the axion, but if two or more $\alpha_i$ coincide there
will be monodromy of infinite order, and potentially interesting
inflationary physics hidden deep in the complex structure moduli space
of the Calabi-Yau threefold $X_3$. In addition, the number of
coincident $\alpha_i$ will determine the index of the monodromy matrix
since it will determine in turn the highest power of the logarithms
appearing in the expansion of the periods. For instance, if two
$\alpha_i$ coincide we expect a very similar behavior to that of the
conifold point. In contrast, if four $\alpha_i$ coincide the behavior
will be similar to that of the large complex structure point.

This can be motivated in a general way using the formalism in
\cite{Greene:2000ci}, that applies whenever the Picard-Fuchs equation
becomes of the form~\eqref{eq:PF}. Consider for example the case
$\alpha_1\neq\alpha_2\neq\alpha_3=\alpha_4$, which we will study in
detail momentarily. According to the results in \cite{Greene:2000ci},
in a certain natural basis of periods (the \emph{Jordan} basis in the
language of \cite{Greene:2000ci}) the monodromy matrix around
$z=\infty$ takes the (exponentiated normal Jordan) form
\begin{equation}
  \label{eq:Jordan}
  T[\infty] = \begin{pmatrix}
    e^{-2\pi i\alpha_1} & 0 & 0 & 0\\
    0 & e^{-2\pi i\alpha_2} & 0 & 0\\
    0 & 0 & e^{-2\pi i\alpha_3} & 0\\
    0 & 0 & 2\pi i\,e^{-2\pi i\alpha_3} & e^{-2\pi i\alpha_3}
  \end{pmatrix}\, .
\end{equation}
Such a monodromy transformation arises from periods with the leading
behavior at large $z$ given by $(z^{-\alpha_1}, z^{-\alpha_2}, z^{-\alpha_3},
z^{-\alpha_3}\log(z))$. The last period is the logarithmic term that
we are after. For very large values of $|z|$ the system will have an
approximate shift symmetry under $z\to e^{i\beta} z$, and we can
identify the approximate flat direction with the phase of $z$.

Similarly, if $\alpha_1=\alpha_2\neq\alpha_3=\alpha_4$ the monodromy
on the Jordan basis becomes
\begin{equation}
  T[\infty] = \begin{pmatrix}
    e^{-2\pi i\alpha_1} & 0 & 0 & 0\\
    2\pi i\,e^{-2\pi i\alpha_1} & e^{-2\pi i\alpha_1} & 0 & 0\\
    0 & 0 & e^{-2\pi i\alpha_3} & 0\\
    0 & 0 & 2\pi i\,e^{-2\pi i\alpha_3} & e^{-2\pi i\alpha_3}
  \end{pmatrix}
\end{equation}
and we expect two independent periods to have $\log$ behavior close to
$|z|=\infty$.

In this way one can obtain the expected $\log$ behavior of a given
geometry close to $z=\infty$. This gives a useful guide in searching
for geometries having axions with different qualitative forms of their
potential. We will now analyze in detail a number of situations where
the above expectation for the behavior of the periods can be checked,
and compute the low energy physics (in the complex structure sector,
ignoring other aspects of the background) for the corresponding flux
compactifications.

\subsection{A symplectic integral basis of periods at large complex
  structure}\label{sec:large}

\begin{figure}[h!]
\centering
\vspace*{4cm}
\begin{picture}(100,100)
\put(-90,0){\includegraphics[width=0.45\textwidth]{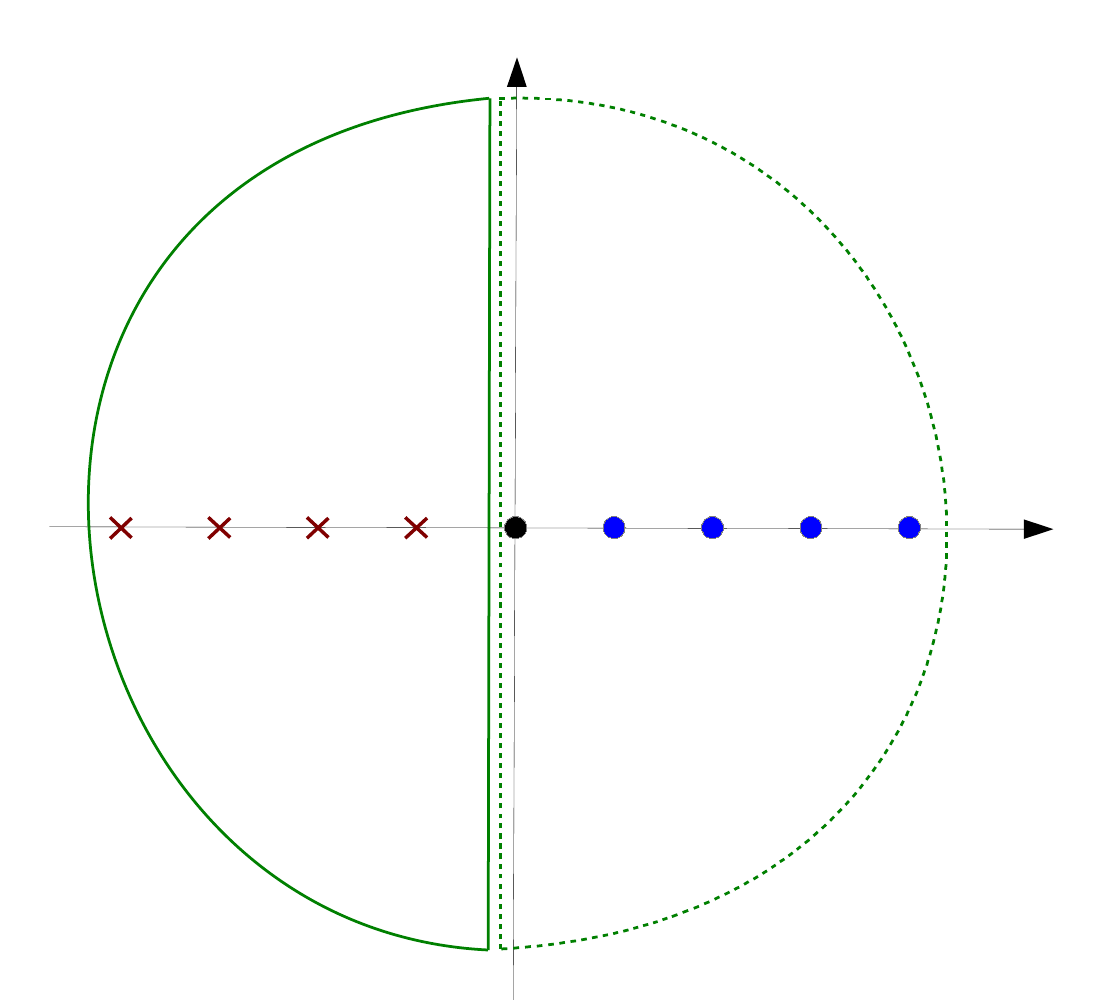}}
\put(80,150){$s$}
\put(-25,95){$-\alpha_1$}
\put(18,95){$1$}
\put(38,95){$2$}
\end{picture}
\begin{minipage}{12cm}
  \caption{Schematic depiction of the $s$-plane including two
    contours, dotted green line and solid green line. The crosses
    indicate poles at $s = -\alpha_i - n$ with $n=\{0,1,2,\ldots \}$,
    while the small spheres indicate poles at $s=\{0,1,2,\ldots
    \}$.} \label{fig:Contours}
\end{minipage}
\end{figure}
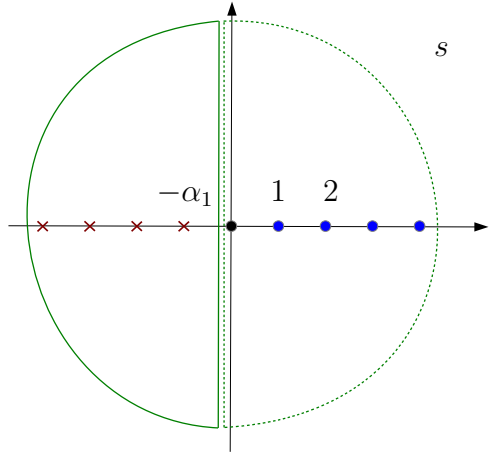

Our first task is to find an integral and symplectic basis of
solutions $\Pi_i$ to equation~\eqref{eq:PF}. Let us start by finding a
basis of solutions, not necessarily integral nor symplectic. Since we
will want to do various analytic continuations, a convenient choice is
to write the basis of solutions in terms of a Mellin-Barnes
representation. A nice way of constructing Mellin-Barnes
representations of the solutions that fits perfectly the task at hand
starts by noticing \cite{Greene:2000ci} that~\eqref{eq:PF} is
precisely a specific type of Meijer G equation. The solutions to such
an equation are aptly named Meijer G functions, and their integral
representation is well known \cite{Olver:2010:NHMF}:
\begin{equation}
  \label{eq:Meijer-g}
  U_j(z) = \frac{1}{(2\pi i)^{j+1}} \int_C\! ds\, c(s) \left(\Gamma(s+1)
    \Gamma(-s)\right)^{j+1} \bigl(z e^{\pi i (j+1)}\bigr)^s
\end{equation}
with \cite{Hosono:1994ax}
\begin{equation}
  \begin{split}
    c(s) & \equiv
    \kappa^s \frac{\prod_{i=1}^4\Gamma(s+\alpha_i)}{\Gamma(s+1)^4\prod_{i=1}^4\Gamma(\alpha_i)}\\
    & = \frac{\prod_{i=1}^k\Gamma(d_is+1)}{\Gamma(s+1)^{n+1}}\, .
  \end{split}
\end{equation}
The integration contour $C$ in~\eqref{eq:Meijer-g} is taken to be a
straight line going from $\varepsilon -i\infty$ to $\varepsilon +
i\infty$, with $\varepsilon$ a small negative number such that
$-\varepsilon<\alpha_i\,\,\forall \alpha_i$ as depicted in figure
\ref{fig:Contours} by the vertical part of the (dotted or continuous)
green contours.  An easy exercise shows that $\cL_\alpha U_j(z)=0$.

Clearly $\cL_\alpha$ is a linear operator, so let us try to find an
integral and symplectic basis of solutions by taking linear
combinations of the $U_j(z)$. The basic idea is the following
\cite{Mayr:2000as}: the periods of $\Omega$ in an integral and
symplectic basis of $H_3(X_3,\bZ)$ can be understood physically as the
masses of a set of $D3$ branes\footnote{We take the convention that a
  $Dp$ brane wraps a $p$-cycle in the Calabi-Yau.} wrapping a set of
special Lagrangian representatives of the basis
elements.\footnote{Whether such representatives actually exist at a
  given point in moduli space depends on considerations of stability,
  see \cite{Aspinwall:2004jr} for a review.  This subtlety is not
  important for the argument here, since the objects we construct will
  exist in the large volume regime of the mirror $Y_3$, and this is
  enough to determine a basis of homology with the desired
  properties.} By mirror symmetry, these branes will be mapped to a
set of $D(2p)$ branes on $Y_3$. The mass of such $D(2p)$ branes can be
calculated reliably in the large volume limit. By matching with the
behavior of the basis of solutions $U_j(z)$ near $z=0$, and making use
of the mirror map, we can fix an integral basis of periods.

We are interested in the expansion of~\eqref{eq:Meijer-g} around $u=0$
keeping only the classical terms (i.e. $\log$ and constant terms, but
no positive powers of $z$). For computing this we can close the
integral to the right, as depicted in figure \ref{fig:Contours} by 
a dotted green line, and compute the residue at $s=0$. The other
poles we pick up inside the contour (coming from $\Gamma(-s)$) at
$s=\{1,2,\ldots\}$ will have a residue going as $z^k$ with $k>0$, so
for the perturbative expansion they can be ignored. In this regime we
have
\begin{equation}
  U_j(z) = -\Res_{s=0} \left\{\frac{1}{(2\pi i)^j}c(s) \left(\Gamma(s+1)
    \Gamma(-s)\right)^{j+1} \bigl(ze^{\pi i (j+1)}\bigr)\right\}\, .
\end{equation}

The rest of the exercise is lengthy but straightforward, one just
needs the expansion of the $\Gamma$ function to sufficiently high
order
\begin{equation}
  \Gamma(-s) = -\left[\frac{1}{s} + \gamma +
    \frac{1}{2}\biggl(\gamma^2 + \frac{\pi^2}{6}\biggr)s +
    \frac{1}{6}\biggl(\gamma^3 + \frac{\gamma \pi ^2}{2} + 2 \zeta(3)\biggr)s^2\right] + \cO(s^3)
\end{equation}
with $\gamma$ the Euler-Mascheroni constant, a few identities relating
to the digamma function $\psi(s) = \partial_s \log(\Gamma(s))$ and its
derivatives with respect to $s$
\begin{equation}
    \psi(1)  = -\gamma \ , \qquad 
    \partial_s\psi(1)  = \frac{\pi^2}{6}\ , \qquad
    \partial_s^2\psi(1)  = -2\zeta(3)\ ,
\end{equation}
and finally some identities relating $c(s)$ to topological data of $Y_3$
\cite{Hosono:1994ax}
\begin{equation}
    c(0)  = 1\ , \quad 
    \partial_sc(0)  = 0\ ,\quad 
    \partial_s^2c(0)  = b\pi^2\ , \quad 
    \partial_s^3c(0)  = \frac{6}{\Kint} \zeta(3)\chi\, ,
\end{equation}
where
\begin{align}
    \chi & = \int_{Y_3} c_3(TY_3)\ , \qquad  \Kint  = \int_{Y_3} D\wedge D \wedge D\ , \qquad 
    b = \frac{1}{3\Kint}\int_{Y_3} c_2(TY_3) \wedge D \nn \\
    \begin{split}
      \label{eq:bhat}
      \hat{b} & = \frac{\Kint}{8}b 
       = \frac{1}{24} \int_{Y_3} c_2(TY_3) \wedge D\, .
    \end{split}
\end{align}
Here $D$ is a divisor of $Y_3$ generating $H^2(Y_3,\bZ)$, coming from
the restriction to $Y_3$ of the hyperplane of the ambient toric space,
and $c_i(TY_3)$ are the Chern classes of the tangent bundle of $Y_3$.

With all this in place, one can immediately compute the periods, for example
\begin{equation}
  \begin{split}
    U_1(z) & = -\Res_{s=0}\left\{\frac{1}{2\pi i}c(s)
      \left(\Gamma(s+1)
        \Gamma(-s)\right)^2 \bigl(ze^{2\pi i}\bigr)\right\}\\
    & = -\Res_{s=0}\left\{\frac{1}{2\pi i}c(s) \Gamma(s+1)^2
      \left(\frac{1}{s^2} +
        2\gamma\frac{1}{s}\right) \bigl(ze^{2\pi i}\bigr)\right\}\\
    & = \frac{1}{2\pi i}\Bigl[-2\gamma \underbrace{\left(c(0)\Gamma(1)^2\right)}_{1}
    - \partial_s\left(c(s)\Gamma(1+s)^2 \left(ze^{2\pi i
          s}\right)\right)|_{s=0}\Bigr]\\
    & = \frac{1}{2\pi i}\Bigl[-2\gamma - \left(2\psi(1) + \log\left(ze^{2\pi
          i}\right)\right)\Bigr]\\
    & = - \frac{\log(z)}{2\pi i} - 1\, .
  \end{split}
\end{equation}
The other basis elements can be expanded similarly, we obtain (up to
$\cO(z)$ terms)
\begin{subequations}
  \label{eq:Meijer-expansion}
  \begin{align}
    \label{eq:Meijer-expansion-a}
    U_0(z) & = 1\\
    U_1(z) & = - \frac{\log(z)}{2\pi i} - 1\\
    U_2(z) & = \frac{1}{2}\left(\frac{\log(z)}{2\pi i}\right)^2 +
    \frac{3}{2}\frac{\log(z)}{2\pi i} +
    \frac{8-b}{8}\\
    \label{eq:Meijer-expansion-d}
    U_3(z) & = -\frac{1}{6}\left(\frac{\log(z)}{2\pi i}\right)^3 -
    \left(\frac{\log(z)}{2\pi i}\right)^2 +
    \frac{\log(z)}{2\pi i}\frac{(3b - 44)}{24} + \frac{(2b - 8)}{8} -
    \frac{\chi}{(2\pi i)^3 \Kint}\zeta(3)\, .
  \end{align}
\end{subequations}
The mirror map can now be constructed as usual by taking a period
linear in $\log(z)$ with the right coefficient. A convenient choice,
given our basis of solutions, is
\begin{equation}
  t  =-\frac{U_1(z)+U_0(z)}{U_0(z)}=\frac{1}{2\pi i}\log(z) + \cO(z)
\end{equation}
where $t$ denotes the complexified K\"ahler modulus of $Y_3$. More
explicitly, the complexified K\"ahler two-form on $Y_3$ is given by $B+iJ= t D$,
where $B$ is the NS-NS B-field and $J$ is the K\"ahler form. 
 
\medskip

Now that we have a basis of periods, and its expansion around the
mirror of the large volume point of $Y_3$, let us construct an integral
basis by matching with the periods of a suitable basis of branes on
$Y_3$. A
natural integral (but not necessarily symplectic) basis of objects in
this regime is given by
\begin{itemize}
\item A $D6$ wrapping $Y_3$.
\item A $D4$ on the divisor $D$.
\item A $D2$ on the curve $C$ dual to $D$ ($C\cdot D=1$).
\item A $D0$ on a point on $Y_3$.
\end{itemize}
Notice that are underspecifying the branes: at the very least we
should describe the bundle on them. Or, more precisely, we should be
specifying an object in the derived category of coherent sheaves. We
will do so momentarily.

We follow the notation and conventions of \cite{Aspinwall:2004jr}, we
refer the reader to this review for the requisite background
material. In the simplest case, a brane is defined by a sheaf with
support on a submanifold $S$ of $Y_3$ (possibly $S=Y_3$). If the
inclusion map is $i\colon S\hookrightarrow Y_3$, such a brane is
described by a sheaf $i_*\cE$. As explained
in~\cite{Aspinwall:2004jr}, to such a brane we can assign a charge
vector
\begin{equation}
  \label{eq:brane-charge}
  \Gamma(i_*\cE) = \ch(i_*\cE)\sqrt{\Td(TY_3)}
\end{equation}
with $\Td(TX)$ the Todd class of the tangent bundle to $Y_3$. The
expression simplifies for sheaves supported on a submanifold $S$, in
this case \cite{Aspinwall:2004jr}
\begin{equation}
  \Gamma(i_*\cE) = PD[S]\wedge \ch(\cE)\wedge
  \sqrt{\frac{\Td(TS)}{\Td(NS)}}
\end{equation}
where $PD[S]$ is the Poincar\'e dual form to $S$.

We are now in a position to identify our integral basis (not yet
symplectic). The $D6$ will be the brane associated to the trivial
sheaf on $Y_3$, i.e. $\cO_{Y_3}$. The $D0$ will be the skyscraper sheaf on a
point. The $D2$ will be $j_*\cO$ with $j\colon C\hookrightarrow Y_3$,
and the $D4$ will be $k_*\cO$, with $k\colon D\hookrightarrow Y_3$.

In the classical limit (i.e. close to $z=0$) the periods for such a
brane wrapped on a manifold $S$ are easily computed
\cite{Aspinwall:2004jr}
\begin{equation}
  \label{eq:B-Z}
  Z(i_*\cO_S) = \int_{Y_3} e^{-tD}\wedge \Gamma(i_*\cO_S) = \int_S
  i^*(e^{-tD})\wedge \sqrt{\frac{\Td(TS)}{\Td(NS)}}
\end{equation}
with $t$ the mirror parameter. Using well known expression for the
Todd class of a bundle $E$
\begin{equation}
  \Td(E) = 1 + \frac{1}{2}c_1 + \frac{1}{12}(c_1^2+c_2) +
  \frac{1}{24}c_1 c_2 + \ldots
\end{equation}
where we have omitted terms vanishing for complex threefolds. We get
the following central charges~\eqref{eq:B-Z} for our integral basis of
branes (here $i_*\cO_p$ is the trivial sheaf supported on the $Dp$
brane)
\begin{subequations}
\begin{align}
  \label{eq:D0-period}
  Z(i_*\cO_0) & = 1\\
  Z(i_*\cO_2) & = -t + 1 - g_C\\
  \label{eq:D4-period}
  Z(i_*\cO_4) & = \frac{\Kint}{2}t^2 - \frac{1}{2} \Kint t +
  \frac{\Kint}{6} + \frac{1}{24} \int_D c_2(TD)\\
  \label{eq:D6-period}
  Z(i_*\cO_6) & = -\frac{\Kint}{6}t^3 - t\hat{b} - \hat{c}\, ,
\end{align}
\end{subequations}
where $\hat b$ and $\Kint$ have been introduced in \eqref{eq:bhat} and
we have encoded in $\hat{c}$ the one-loop term $\hat c =
\frac{\zeta(3)\chi}{(2\pi i)^3}$.\footnote{Let us point out that the
  expression~\eqref{eq:brane-charge} should actually use the so-called
  $\Gamma$-class instead of the Todd class
  \cite{Libgober,Iritani1,KKP,Iritani2,Halverson:2013qca,Hori:2013ika,Kim:2013ola}.
  The correction is precisely the $\hat{c}$ term which we have simply
  added by hand to~\eqref{eq:D6-period}, the rest of the central
  charges~\eqref{eq:D0-period}-\eqref{eq:D4-period} are unaffected.} The integer $g_C$ is the
genus of $C$.

We see that the qualitative form of this set of central charges
matches nicely with the structure of the periods in $X_3$ near $z=0$
found in~\eqref{eq:Meijer-expansion}. This basis is clearly made of
integral objects, but it is not necessarily symplectic. In order to
achieve this, we will take integral linear combinations of these
elements in such a way that the chiral spectrum in the new basis is of
the symplectic form. By the Hirzebruch-Riemann-Roch theorem and the
expression~\eqref{eq:brane-charge} the net chiral spectrum between
branes $A$ and $B$ is counted by \cite{Aspinwall:2004jr}
\begin{equation}
  \dsz{A}{B} \equiv \int_{Y_3} \ch(i_*\cE^\vee_A) \ch(i_*\cE_B)\Td(TY_3) =
  \int_{Y_3} \Gamma(A)^\vee \wedge \Gamma(B)
\end{equation}
with $\Gamma(A)^\vee$ means that the even components of the polyform
$\Gamma(A)$ should be changed sign. (Notice that this product, being
mirror to an intersection product between $D3$ branes, is
antisymmetric as it should, and it is always an integer.) We obtain:
\begin{align}
  \dsz{D6}{D0} & = 1\ , &&
  \dsz{D6}{D2}  = 1-g_C   \\
  \dsz{D6}{D4} & = \frac{\Kint}{6} + \frac{1}{24}\int_Dc_2(TD) +
  \hat{b}\ , &&   \dsz{D4}{D2}  = -1\ ,  \nn \\
  \dsz{D4}{D0} & = 0 \ , &&  \dsz{D2}{D0}  = 0\ . \nn 
\end{align}
We see that the basis is almost symplectic. We can make it completely
symplectic by choosing as a basis $\cS = \{D0, -D2+(1-g_C)D0, -D4 +
\dsz{D6}{D4}D0, -D6 \}$. The classical periods in this basis, which is
now the integral symplectic basis that we are after, are
\begin{equation}
\label{PiSlarge}
  \Pi^\cS = \begin{pmatrix}
    1\\
    t\\
    -\frac{\Kint}{2}t^2 + \frac{\Kint}{2}t + \hat{b}\\
    \frac{\Kint}{6}t^3 + t\hat{b} + \hat{c}
  \end{pmatrix}
\end{equation}
This agrees with the expression in \cite{Huang:2006hq}.

We can now match this perturbative behavior with the expansion of the
$U_i(z)$ basis found in~\eqref{eq:Meijer-expansion}, with the result
that we can write the expression for the periods in the integral
symplectic basis in terms of the exact solutions $U_i(z)$. We have
that $\Pi^\cS_i = \Xi_{ij}U_j(z)$, with
\begin{equation}
  \label{eq:base-change}
  \Xi_{ij} = \begin{pmatrix}
    1 & 0 & 0 & 0 \\
    -1 & -1 & 0 & 0 \\
    -\Kint & -2 \Kint & -\Kint & 0 \\
    -\left(\frac{\Kint b}{4} + \frac{\Kint}{6}\right) &\,\,
    -\left(\frac{\Kint b}{4} + \frac{7\Kint}{6}\right)\,\, & -2 \Kint & -\Kint
    \end{pmatrix}\, .
\end{equation}

This is all the information we need in order to start exploring the
physics away from $z=0$.  
But before turning to this problem let us comment about the four-dimensional $\cN=1$ effective theory that we get near the large complex structure point. Inserting the periods of \eqref{PiSlarge} into \eqref{Kinperiods} and \eqref{Winperiods} we get the following K\"ahler potential and the $\cN=1$ superpotential
 \begin{align} 
    K_{\rm lcs}&= - \log\big[ i (\tau-\bar \tau) \big]- \log \left[i\left(-\frac{1}{6} \cK (t-\bar{t})^3+ 2 \hat c\right)+\dots\right] \ ,  \\[.1cm]
    W_{\rm lcs} &= \frac{1}{6} \cN_4 \Kint t^3-\frac{1}{2}\cN_3 \Kint t^2+\left(\cN_4\hat b+\frac{1}{2}\cN_3 \Kint+\cN_2\right)t+\left(\cN_1-\cN_4\hat{c}+\cN_3\hat{b}\right) \ ,  \nn
\end{align}
where $\cN_i$ was defined in \eqref{Winperiods} and correspond to the
flux quanta. We can see that $K$ preserves an approximate continuous
shift symmetry for the real part of the field $t$, which would be
broken by non-perturbative corrections to $K$ as we move away from the
special point $z=0$ ($\text{Im}(t)\rightarrow \infty$). These
non-perturbative contributions of $\cO(e^{2\pi i t})$ come from the
$\cO(z)$ terms in the expansion of the periods in
\eqref{eq:Meijer-expansion-a}-\eqref{eq:Meijer-expansion-d}, and break
the continuous symmetry to a discrete periodicity. This discrete
symmetry remains exact since it belongs to the discrete monodromy
symmetries preserved by $K$ near the large complex structure
point. However if we are close enough to the special point these
corrections are negligible and we recover an approximate continuous
shift symmetry for the field. This fact has been used in several
occasions to realize large field inflation models within the complex
structure moduli space of Calabi-Yau manifolds
\cite{Marchesano:2014mla,Hebecker:2014eua,
  Hayashi:2014aua,Blumenhagen:2014nba,Hebecker:2014kva}. In these
models the shift symmetry is broken by the flux induced superpotential
derived above, giving rise to the usual polynomial-type potentials.

However it is particularly simple, and physically
relevant, to explore the behavior of the periods close to
$z=\infty$, where new types of scalar potentials arise. We turn to this problem next.

\subsection{Continuation to $z=\infty$: generalities}

We just obtained the expansion around $z=0$ of the
periods~\eqref{eq:Meijer-g} by closing the integration contour in
figure~\ref{fig:Contours} to the right. This does not modify the value
of the integral when $|u|<1$, and in this regime the sum over the
residues gives rise to a convergent series. Neither of these
statements is true when $|u|>1$, but a similar idea applies: when
$|u|>1$ one can close the contour \emph{to the left} without modifying
the integral, and the resulting sum over residues gives a convergent
series. This contour is depicted in figure \ref{fig:Contours} by a
solid green line,

The singular behavior of the integrand comes, as before, from the
poles of the $\Gamma$ functions at non-positive integral
values. Recall from~\eqref{eq:Meijer-g} that the integrand for $U_j$
has a expression of the form (focusing on the $\Gamma$ function part)
\begin{equation}
  U_j(z) \sim \int\!ds\, \frac{\prod_{i=1}^4\Gamma(s+\alpha_i)}{\Gamma(s+1)^{3-j}}\ldots
\end{equation}
It is clear that the poles for $\text{Re}(s)<0$ come only from the
numerator, when $s = -\alpha_i - n$, with $n\in\{0,1,\ldots\}$.

Furthermore, now we see clearly the origin of our claim before that
the qualitative behavior of the low energy physics close to $z=\infty$
will depend on the structure of the $\alpha_i$: if they all different
all the poles picked up by the integral will be single poles, and
there are no $\log(z)$ factors in the periods. If they are double
poles, the corresponding terms will be linear in $\log$s, etc. For
having interesting monodromies we will want to have at least two
coincident $\alpha_i$. We will now analyze in detail some interesting
scenarios.

We focus on the case of having double poles. Triple and quadrupole
poles (corresponding to $\alpha_1=\alpha_2=\alpha_3\neq\alpha_4$ and
$\alpha_1=\alpha_2=\alpha_3=\alpha_4$) are analyzed similarly. The
particular case of all $\alpha_i$ equal is conceptually interesting:
in this case one would expect the behavior close to $z=0$ to be
qualitatively similar to that at $z=\infty$. Study of concrete
examples supports this expectation: the canonical example of this
situation, $\bP^7[2,2,2,2]$ has a $z\leftrightarrow 1/z$ symmetry in
moduli space, as pointed out in \cite{Berglund:1993ax}.

\subsection{$\alpha_1\neq\alpha_2\neq \alpha_3 = \alpha_4$}

Let us consider the minimal non-trivial case in which two eigenvalues
coincide,
\begin{equation}
\alpha_1\neq \alpha_2\neq \alpha_3=\alpha_4\, .
\end{equation}
In this case we have two contributions to the periods coming from the
single poles ($s=-\alpha_1$ and $s=-\alpha_2$) and from the double
pole ($s=-\alpha_3)$,
\begin{equation}
U_j(z)=U_j^s(z)+U_j^d(z)
\end{equation}
For instance, the contribution from the single poles at $s=-\alpha_1$
and $s=-\alpha_2$ is given by ($k=1,2$)
\begin{equation}
 U_j^s(z) =\Res\left[\frac{1}{(2\pi i)^j} c(s) \left(\Gamma(s+1)
    \Gamma(-s)\right)^{j+1} \bigl(ze^{\pi i (j+1)}\bigr)^s\right]_{s=-\alpha_k}
\end{equation}
leading to
\begin{equation}
  U_j^s(z)=\frac{A_k}{(2\pi i)^j}\left(\frac{\sin(\pi\alpha_k)}{\pi}\right)^{3-j}e^{-\pi
    i(j+1)\alpha_k}(\kappa z)^{-\alpha_k}
\end{equation}
with
\begin{equation}
  \label{eq:Ak}
  A_k=\frac{\Gamma(\alpha_k)^4\Pi_{i\neq
      k}\Gamma(\alpha_i-\alpha_k)}{\Pi_{i=1}^4\Gamma(\alpha_i)}\, .
\end{equation}
To compute the residue we have used the reflection formula
\begin{equation}
\Gamma(1-s)\Gamma(s)=\frac{\pi}{\sin(\pi s)}\ .
\end{equation}
The contribution from the double pole at $s=-\alpha_3$ is given by
\begin{equation}
  \begin{split}
    U_j^d(z) & =\Res \left[ \frac{1}{(2\pi i)^j}\frac{\Gamma(s+\alpha_1)\Gamma(s+\alpha_2)}{\Pi_{i=1}^4\Gamma(\alpha_i)\Gamma(s+1)^4} \left(\frac{-2\gamma}{s+\alpha_3}+\frac{1}{(s+\alpha_3)^2}\right) \right. \\[.1cm]
    & \phantom{=}\qquad\quad \ \times \left. \left(\Gamma(s+1)
        \Gamma(-s)\right)^{j+1} \bigl(ze^{\pi i (j+1)}\bigr)^s\right]_{s=-\alpha_3}\ ,
  \end{split}
\end{equation}
which yields
\begin{equation}
  \label{eq:Uj-double-pole}
  \begin{split}
    U_j^d(z)&=\frac{A_3}{(2\pi i)^j}\left(\frac{\sin(\pi\alpha_3)}{\pi}\right)^{3-j}e^{-\pi
      i(j+1)\alpha_3}(\kappa z)^{-\alpha_3} \\
  & \qquad \qquad \times \left(B-(3-j)\pi
    \cot(\pi\alpha_3)+(j+1)i\pi+\log(\kappa z)\right)
\end{split}
\end{equation}
with
\begin{equation}
  B=\psi(\alpha_1-\alpha_3)+\psi(\alpha_2-\alpha_3)-4\psi(\alpha_3)-2\gamma\ ,
\end{equation}
and $A_3$ as in~\eqref{eq:Ak}. Here we have also used the identity
\begin{equation}
\psi(1-s)-\psi(s)=\pi\cot(\pi s)\, .
\end{equation}
Notice that unlike the single poles, the double pole induces a
logarithmic term in the periods.

We can now write the leading expansion of the integral symplectic
basis using the relation~\eqref{eq:base-change}. For concreteness let
us consider a particular example, given by $\bP^5[2,4]$. In this
geometry we have $\Kint=8$, $\hat{b}=b=\frac{7}{3}$, $\chi=-176$,
$\kappa=2^{10}$, and $\alpha_i=\{\frac{1}{4}, \frac{3}{4},
\frac{1}{2}, \frac{1}{2}\}$. The periods in the Meijer basis read
\begin{equation}
U_j(z)=U_j^s(z)+U_j^d(z)
\end{equation}
where
\begin{align}
  \label{eq:Meijer-large-z}
    U_j^s(z)&=\frac{1}{(2\pi
      i)^j}\frac{\Gamma(1/4)^6}{(2\pi^3)^{1/2}}(\sqrt{2}\pi)^{j-3}e^{-\pi
      i(j+1)/4}(\kappa z)^{-1/4}+\mathcal{O}(z^{-3/4})\, , \\
    U_j^d(z)&=-\frac{1}{(2\pi i)^j}\frac{4\pi}{\pi^{3-j}}e^{-\pi
      i(j+1)/2}(\kappa z)^{-1/2}\left(2\log(2)+4+(j+1)i\pi+\log(\kappa
      z)\right) + \cO(z^{-3/4})\, .
\end{align}
To compute the second contribution we have evaluated the Gamma
functions at the particular values of $\alpha_i$ corresponding to
$\bP^5[2,4]$, obtaining
\begin{gather}
A_3=\frac{\Gamma(1/2)^2\Gamma(-1/4)\Gamma(1/4)}{\Gamma(1/4)\Gamma(3/4)}=-4\pi\\
B=\psi(-1/4)+\psi(1/4)-4\psi(1/2)-2\gamma=2\log(2)+4\, .
\end{gather}

Using the change of basis~\eqref{eq:base-change} this gives rise to
the following expansion of the integral symplectic periods around the
small complex structure point
\begin{equation}
  \Pi^\cS = \frac{4i}{\pi^2}u^{-1/2}\
  \begin{pmatrix} B + i \pi + \log(u)\\
    \tfrac12 (-B - \log(u))\\
    \tfrac14 (-B\Kint - 8 \pi - \Kint \log(u))\\
    -\tfrac{1}{24} (24 \hat{b} - \Kint) (B + \log(u))
  \end{pmatrix}+\frac{\Gamma(1/4)^6}{4\pi^4\sqrt{\pi}}u^{-1/4}\begin{pmatrix}
-e^{3\pi i/4}\\  \frac{i}{\sqrt{2}}\\ \frac{1}{2} \Kint e^{i\pi/4}\\ \frac{i}{3\sqrt{2}}(6\hat{b}-\Kint)\end{pmatrix}+\dots
\end{equation}
(recall $u\equiv\kappa z$) which implies a K\"ahler potential given by
\begin{equation}
  \begin{split}
  K& =-\log\left(i\Pi^{\cS\dag}\Sigma\Pi^\cS\right)\\
  & =-\log\left(\frac{1}{48\pi^6|u|}(\pi^32\,\text{Re}(a_1
    \log(u))-\pi^22\,\text{Re}(a_2 u^{1/4})+a_3|u|^{1/2}+a_4)+\ldots\right)
\end{split}
\end{equation}
where
\begin{align}
a_1&=32(24i+24\hat{b}-\Kint)=1536-768i\\
a_2&=A_1(2+2i)(4-4i)(6i+6b-\Kint)=A_1(96+96i)\\
a_3&=A_1^2(12\hat{b}-5\Kint)=-12A_1\\
a_4&=64B(24b-\Kint)=3072B
\end{align}
and $A_1=\frac{\Gamma(1/4)^6}{(2\pi^3)^{1/2}}$. Notice that the
leading term when $z\rightarrow \infty$ is of order
$\cO{(|u|^{-1}log(u))}$ while the corrections go as
$|u|^{-1}u^{\alpha_i}$. The cross term $\cO{(u^{1/4}log(u))}$ vanishes
since the coefficient is proportional to $24\hat{b}-7 \Kint=0$.  Here
the coefficients $a_i$ are complex combinations of the topological
intersection numbers so in general will be complex, implying a
quadratic dependence of $e^{-K}$ on the phase of $u$ (unlike the
sinusoidal modulations coming from the polynomial corrections
$\cO{(u^{\alpha_i})}$). However this term is suppressed by $|u|^{-1}$
so can be neglected if we are close enough to the special point (as
well as the subleading corrections that are not proportional to
$\log(z)$) such that we recover an approximate shift symmetry near the
special point $u=\infty$.

All the integral symplectic periods show a behavior on the coordinate
$u$ given by $\cO{(u^{-1/2}(\log(u)+\ldots))}+\cO{(u^{-1/4})}$. Hence
schematically the superpotential will be given by
\begin{equation}
W=N_{eff}u^{-1/2}(\log(u)+\ldots)+N'_{eff}u^{-1/4}+\ldots
\end{equation}
with $N_{eff},N'_{eff}$ some effective coefficients depending also on
the NSNS and RR fluxes. This implies a scalar potential of the following form
\beq
V=V_0(|u|)+f(|u|)cos(\alpha_3-\alpha_1)\phi+\dots
\eeq
with $\phi$ (the phase of $u$) being the natural candidate for the axion. Notice that $V_0(|u|)$ and $f(|u|)$ are functions only on the modulus $|u|$ so the leading term in the axion is a cosine term. This behaviour will also appear near the conifold point as we will see in section \ref{sec:conifold}. The difference is that here the effective decay constant also depends on the structure of the $\alpha_i$ and there might be an enhancement given by $(\alpha_3-\alpha_1)^{-1}$ (a factor 4 in the case of $\bP^5[2,4]$).

\medskip

Let us come back momentarily to the general discussion in
section~\ref{sec:monodromies}, and check that the generic expectations
in that section are fulfilled in this example. Clearly, taking
appropriate linear combinations of~\eqref{eq:Meijer-large-z} one can
construct a basis of periods with the leading behavior $(z^{-1/4},
z^{-3/4}, z^{-1/2}, \log(z)z^{-1/2})$. The monodromy behavior of this
basis under $z\to e^{2\pi i}z$ is given by~\eqref{eq:Jordan}. Taking
the monodromy in this Jordan basis, and changing back to our
basis~\eqref{eq:Meijer-g}, we obtain a monodromy matrix for
$\bP^5[2,4]$ given by
\begin{equation}
  T^U[\infty] = \begin{pmatrix}
    1 & 6 & 8 & 8 \\
    -1 & -5 & -8 & -8 \\
    1 & 5 & 9 & 8 \\
    -1 & -5 & -9 & -7
  \end{pmatrix}
\end{equation}
which in the integral symplectic basis becomes
\begin{equation}
  T^\cS[\infty] = \begin{pmatrix}
    -5 & -8 & 1 & -1 \\
    1 & 1 & 0 & 0 \\
    0 & -8 & 1 & 0 \\
    6 & 8 & -1 & 1
  \end{pmatrix}\, .
\end{equation}
It is easy to check that $(T[\infty]^4 - I)^2=0$.

\subsection{$\alpha_1=\alpha_2\neq \alpha_3=\alpha_4$}

Let us consider the case in which $\alpha_1=\alpha_2\neq
\alpha_3=\alpha_4$. The poles for $\Re(s)<0$ are located at
$s=-\alpha_1-n$ and $s=-\alpha_3-n$ with $n\in\{0,1,\ldots\}$. Unlike
the previous case, now all the poles are double. Integration around
these poles gives a contribution of the exact
form~\eqref{eq:Uj-double-pole} and another also of the
form~\eqref{eq:Uj-double-pole} with $\alpha_3\to \alpha_1$.

One example manifold with these properties is $\bP^5[3,3]$, with
$\alpha_i=\{\frac{1}{3},\frac{1}{3}, \frac{2}{3}, \frac{2}{3}\}$,
$\Kint=9$, $b=2$, $\hat{b}=\frac{9}{4}$, $\kappa=3^6$ and $\chi=-144$.
The parameters $A_k,B$ become
\begin{align}
&A_1\begin{gathered}[t]=\frac{\Gamma(1/3)^2\Gamma(1/3)^2}{\Gamma(2/3)}^2=\frac{3}{4\pi^2}\Gamma(1/3)^6\end{gathered}\ ,
&
&A_2\begin{gathered}[t]=\frac{\Gamma(2/3)^2\Gamma(-1/3)^2}{\Gamma(1/3)}^2
  =\frac{12\pi^2}{A_1}\ ,\end{gathered} \nn \\[.1cm]
&B_1=-2\psi(1/3)-2\gamma\ , &  &B_2=B_1+6-\frac{2\pi}{\sqrt{3}}\ ,
\end{align}
and the K\"ahler potential at leading order is given by
\begin{multline}
K=-\log\left(-\frac{9}{128A_1^2\pi^6|u|^{4/3}}\left(A_1^46\sqrt{3}(1-i\sqrt{3})|u^{1/3}log(u)|^2+\right.\right.\\
\left.\left.+72\sqrt{3}\pi^2A_1\log(u)|^2((u^{1/3}-\bar{u}^{1/3})-i\sqrt{3}(u^{1/3}+\bar{u}^{1/3}))\right)+\dots \right)\, ,
\end{multline}
Note that the leading term on the K\"ahler
potential goes as $\cO(|u|^{-2/3}\log|u|)$ so we again recover an
approximate shift symmetry for the phase of $u$ in the small complex
structure limit $u\rightarrow\infty$. This shift symmetry would be
broken by a flux-induced superpotential given by
\begin{equation}
W=N_{eff}u^{-1/3}(\log(u)+\dots)+N'_{eff}u^{-2/3}(\log(u)+\dots)+\dots
\end{equation}

In conclusion, when we have one or two pairs of coincident roots
$\alpha_i$, the K\"ahler potential near $u=\infty$ goes as
\begin{equation}
K=-\log\left(\cO(|u^{-\alpha_1}|^2\log|u|) \right)
\end{equation}
where $\alpha_1$ would be the lowest repeated root. In this limit we
recover a shift symmetry for the phase of $u$ which is perturbatively
broken by a superpotential taking the form
\begin{equation}
  W=\sum_iN_{eff_i}u^{-\alpha_i}(\log(u)+\ldots)+\ldots
\end{equation}
with the index $i$ running over the double repeated roots.

\subsection{The conifold point}
\label{sec:conifold}

All the equations of the form (\ref{eq:PF}) have a conifold-like
singularity at $u=1$ around which the monodromy matrix satisfies
$(T[1]-1)^2=0$. Unfortunately the Picard-Fuchs differential equations
around this point do no take the form of a Meijer G equation, so the
method followed in the previous cases does not apply. In
\cite{ChenYangYui} it was proven that (for the manifolds under study
here) there is always a basis of periods in which the monodromy matrix
around the conifold point takes the form
\begin{equation}
T[1]=\left(\begin{array}{cccc}1&0&0&0\\0&1&0&0\\0&0&1&0\\0&-\cK&0&1\end{array}\right)
\label{Tconifold}
\end{equation}
This basis is based on three power series solutions and one
logarithmic solution for the Picard-Fuchs equation.  While the
expansion of the periods has been extensively studied for the case of
the quintic hypersurface in $\bP^4$ \cite{Candelas:1990rm}, there is
no such study for general one-parameter Calabi-Yau threefolds. It is
beyond the scope of this paper to give the quantitative and general
behavior for the periods in any one-parameter threefold, so we will
restrict our analysis to the quintic. However since the qualitative
form of the periods is general for any one-parameter Calabi-Yau
manifold with Picard-Fuchs equation given by~\eqref{eq:PF}, we expect
similar behavior in other manifolds in the class we study.

Let us consider then the mirror of the quintic manifold $\bP^4[5]$. An
integral symplectic basis of periods which undergo the monodromy given
by \eqref{Tconifold} when circling around the conifold is given
by~\cite{Candelas:1990rm,Huang:2006hq}
\begin{equation}
\Pi=\omega_0(z)\left(\begin{array}{c}
    1\\t_c\\\frac{\cK}{2}\frac{1}{2\pi i}t_c^2\\-\frac{\cK}{2\pi i}t_c
    \log t_c-\frac{\cK}{2}\frac{1}{2\pi i}t_c\end{array}\right)\, .
\end{equation}
Recall that $\cK=5$ for the quintic. The coordinate $t_c$ can be written in terms of the $u$ coordinate appearing in (\ref{eq:PF}) as
\begin{equation}
t_c=\delta+\dots\ ,\ \text{with}\ \delta=\frac{u-1}{u}
\end{equation}
Hence the conifold point is located at $\delta=0$ and circling a complete period around this point is equivalent to transform $t_c\rightarrow t_ce^{2\pi i}$. Using the above formula we get the following K\"ahler potential
\begin{equation}
K=-\log\left[\frac{-\cK}{\pi}|t_c|^2 \log|t_c|+\frac12(t_c-\bar t_c)^2+\dots\right]
\label{K1}
\end{equation}
The K\"ahler potential is invariant under the discrete transformation
\begin{equation}
\theta\rightarrow \theta+1
\end{equation}
where $\theta$ is the phase of $t_c$, ie. $t_c=|t_c|e^{2\pi i \theta}$.
This exact discrete symmetry is inherited from the monodromy transformation which leaves the symplectic form invariant, and can be promoted to an approximate shift symmetry if we are close enough to the special point.

The addition of fluxes induces a superpotential given by
\begin{equation}
W=-\cN_4 \cK t_c \log(t_c)+\frac{1}{2}\cN_3 \cK t_c^2+\big(\cN_2- \frac{1}{2} \cN_4 \cK\big)t_c+\cN_1
\label{Wz}
\end{equation}
where $\cN_i$ denote the flux quanta introduced in \eqref{Winperiods}. 
By using the supergravity formulae we obtain the following scalar potential,
\begin{equation}
V=V_0(|t_c|)-\cN_2\cN_3\frac{\pi}{\cK}(\log |t_c|^2) |t_c| \cos(\theta)+\dots
\label{pot_conifold3}
\end{equation}
which corresponds to a cosine-type potential for the axion $\theta$ (phase of $t_c$). While a detailed analysis of the scalar potential is beyond the scope of this paper, we can already remark that this
point of the moduli space could be a good candidate to construct a
model of natural inflation. 

\section{Elliptic fibrations and Mordell-Weil inflation}

\label{special_MW}

We now study a simple two-parameter setup in which inflation can
appear beyond the large complex structure point, namely mirrors for
genus one fibrations.\footnote{These are fibrations where the fiber
  has the topology of a $T^2$. Typically in string theory model
  building these appear as elliptic fibrations, which implies the
  existence of a section, although this condition can be relaxed
  (which can be interesting for model building purposes
  \cite{Braun:2014oya,Morrison:2014era,Anderson:2014yva,Garcia-Etxebarria:2014qua,Mayrhofer:2014haa,Mayrhofer:2014laa})
  and the results here still apply.} We will focus on the large base
limit, which effectively turns the problem into a one-dimensional
problem, and show how infinite order monodromies can arise close to
the small fiber limit. This observation was already made
in~\cite{Klemm:2012sx}, and some of the relevant technical details
were presented there. Technically the analysis is fairly similar to
the one presented in the last section, so we will content ourselves
with outlining some of the most relevant points of the system, without
going into much detail.

A $T^2$ can be realized in a number of different ways. The most
conventional is as a degree 6 hypersurface on a $\bP^{231}$, but other
realizations are as a $\bP^{112}[4]$ hypersurface, a $\bP^{2}[3]$
hypersurface or a $\bP^3[2,2]$ complete intersection. Other
realizations exist, see for example \cite{Braun:2013nqa,Klevers:2014bqa} for 
recent discussions, but we will focus on the ones just presented. They are
commonly known in the literature as $E_8$, $E_7$, $E_6$ and $D_5$
models, respectively.

Using the results in \cite{Hosono:1994ax} one quickly obtains that the
Picard-Fuchs equation for the $T^2$ in these realizations is given by
\begin{equation}
  \label{eq:PF-T^2}
  \cL = \theta^2 - z \bigl(\theta + \alpha_1\bigr)\bigl(\theta +
  \alpha_2\bigr)
\end{equation}
with $\theta$ the logarithmic derivative on the fiber coordinate $z$,
$\alpha_i$ depending on the type of the fiber realization. Concretely,
for $E_8$ we have $\alpha_i=\bigl\{\tfrac{1}{6}, \tfrac{5}{6}\bigr\}$,
for $E_7$ we have $\bigl\{\tfrac{1}{4}, \tfrac{3}{4}\bigr\}$, for
$E_6$ $\bigl\{\tfrac{1}{3}, \tfrac{2}{3}\bigr\}$ and finally for $D_5$
one has $\bigl\{\tfrac{1}{2}, \tfrac{1}{2}\bigr\}$.

The expansion of the periods around large complex structure
(i.e. small $z$) will have a period going as a regular power series in
$z$ and another going as $\log(z)$. Interestingly, we see a rather
marked difference when we analytically continue to large $z$, using
the techniques reviewed in section~\ref{special_oneparameter}: for the
$E_{\{8,7,6\}}$ realizations of the fiber the expansion around
$z=\infty$ will be regular (all the poles are simple poles, since
$\alpha_1\neq\alpha_2$), but for $D_5$ we will have a logarithmic
solution, since $\alpha_1=\alpha_2$, and the poles will be
double. Equivalently, the monodromy matrices around $z=\infty$ will be
of finite order ($T[\infty]^k = I$ for some positive integer $k$) for
$E_{\{8,7,6\}}$, but will be of infinite order for $D_5$. An explicit
computation of periods confirms this \cite{Klemm:2012sx}: if we take
$a=\{1,2,3,4\}$ for $E_8$, $E_7$, $E_6$ or $D_5$ respectively, we have
that the monodromy of the periods around $z=\infty$ is given in a
certain natural basis by
\begin{equation}
  \label{eq:fiber-monodromy}
  T[\infty]=\begin{pmatrix}
    1-a & -1\\
    a & 1
  \end{pmatrix}
\end{equation}
which in a Jordan basis (recall the discussion in
section~\ref{sec:monodromies}) is
\begin{equation}
  T[\infty] = \begin{pmatrix}
    e^{-2\pi i \alpha_1} & 0\\
    0 & e^{-2\pi i \alpha_2}
  \end{pmatrix}
\end{equation}
for $\alpha_1\neq \alpha_2$, and
\begin{equation}
  T[\infty] = \begin{pmatrix}
    e^{-2\pi i \alpha_1} & 0\\
    2\pi i e^{-2\pi i\alpha_1} & e^{-2\pi i \alpha_1}
  \end{pmatrix}
\end{equation}
if $\alpha_1=\alpha_2$.

Let us now fiber this over a base, so we obtain a Calabi-Yau
threefold. For concreteness, let us choose the base to be $\bP^2$, and
the fibration to have global sections. The fibration structure in this
case was given in \cite{Andreas:1999ty}. The Picard-Fuchs equations for
the whole system are given by \cite{Alim:2012ss,Klemm:2012sx}
\begin{equation}
  \label{eq:PF-elliptic}
  \begin{split}
    \cL_1 & = \theta_1(\theta_1 - 3\theta_2) - z_1(\theta_1 +
    \alpha_1)(\theta_1 + \alpha_2)\\
    \cL_2 & = \theta_2^3 - z_2(\theta_1 - 3\theta_2)(\theta_1 -
    3\theta_2 - 1)(\theta_1 - 3\theta_2 - 2)\, .
  \end{split}
\end{equation}
Here $z_1$ is the fiber coordinate, $z_2$ the base, and $\alpha_i$
parameterize the type of the fiber, just as before. In this context
the parameter $a$ in~\eqref{eq:fiber-monodromy} has a natural
interpretation as the number of sections of the compactification
\cite{Andreas:1999ty}.

The limit of large volume of the base corresponds formally to setting
$\theta_2=0=z_2$. Then the Picard-Fuchs
operator~\eqref{eq:PF-elliptic} reduces to the Picard-Fuchs operator
on the torus fiber~\eqref{eq:PF-T^2}, and the conclusions that we
obtained there carry over straightforwardly. In particular, there is a
marked difference between the $E_{\{8,7,6\}}$ fibration types and the
$D_5$ fibration type, or more intrinsically a marked physical
difference depending on the number of sections the fibration
possesses.

As advanced in section~\ref{threefoldresults}, the number of sections
enters in a rather interesting way in a number of important physical
features of string compactifications. Here we just saw a new
interesting feature of the Mordell-Weil group of the mirror of a given
compactification (assuming that the mirror is elliptically fibered):
if its rank is high enough, we have a natural candidate for the axion
in the limit where the fiber of the mirror is very small.

\section{Conclusions}

In this paper we initiated a systematic study of the global structure
of the complex structure moduli space of Calabi-Yau threefolds and
fourfolds with application to building inflationary models in $\cN=1$
flux compactifications. We focused our analysis on the K\"ahler
potential $K_{\rm cs}= - \log \int \Omega \wedge \bar \Omega$ and flux
superpotential $W_{\rm cs} = \int \Omega \wedge G$, where $G$ is a
three-form of four-form flux. Globally $K_{\rm cs}$, or rather the
corresponding K\"ahler metric, has a discrete symmetry group known as
the monodromy group $G_{\rm mon}$. Specific monodromies arise when
encircling certain special points in the complex structure moduli
space. We showed that close to some of these points, namely the ones
with infinite-order monodromies, an approximate continuous symmetry in
the K\"ahler potential $K_{\rm cs}$ emerges. The corresponding degree
of freedom in the low-energy effective action is an axion-like complex
structure modulus with an approximate shift-symmetry in the K\"ahler
potential. Both the discrete and continuous symmetry are broken by a
sufficiently general flux superpotential. Expanding $W_{\rm cs}$
around the points with an axion one can then evaluate the scalar
potential. We have summarized our findings about the different
K\"ahler potentials and superpotentials in section
\ref{threefoldresults}.

Let us comment on some interesting relations of our set-up to other
inflationary models. Firstly, we pointed out that close to conifold
points in the complex structure moduli space of Calabi-Yau threefolds
there is a natural axionic degree of freedom. The continuous symmetry
is then broken to a discrete symmetry by a flux superpotential and a
scalar potential of cosine-type appears. It is well-known
\cite{Vafa:2000wi,Cachazo:2001jy,Dijkgraaf:2002dh,Heckman:2007ub,Aganagic:2007py}
that in some cases there is a dual description of this set-up obtained
by performing a conifold transition to a resolved configuration in
which the singularity is replaced by a $\mathbb{P}^1$.  In this dual
setting, the axionic degree of freedom is a scalar that complexifies
the resolution volume of the conifold singularity. The three-form flux
translates after transition into the fact that there is a stack of
$D5$-branes on the resolving $\bP^1$ with a superpotential from a
gaugino condensate. In our orientifold set-up an $\cN=1$ version of
this duality could be applicable. In other words, using conifold
singularities and three-form fluxes one can aim to engineer aligned
axion models or N-flation models.

We also briefly considered F-theory compactifications on Calabi-Yau
fourfolds. These are manifestly $\cN=1$ configurations and the
dualities arising at different points in the complex structure moduli
space have been much less studied. One interesting recent observation
has been made for geometric transitions with conifold curves in the
fourfold
\cite{Intriligator:2012ue,Garcia-Etxebarria:2014qua}. Performing such
a transition and identifying a complex structure modulus $z
\rightarrow e^{iG}$, where $G$ is a linear combination of R-R and
NS-NS two-form degrees of freedom, one should be able to obtain the
inflationary scenarios suggested in \cite{Grimm:2007hs,Grimm:2014vva}.
It should be interesting to pursue this further.

Let us stress that despite the absence of a detailed understanding of
$\cN=1$ dualities, the period computations can be performed using the
techniques developed in
\cite{Mayr:1996sh,Klemm:1996ts,Alim:2009bx,Grimm:2009ef,Bizet:2014uua},
and the K\"ahler potential and superpotential can be computed at
various points in the complex structure moduli space. It would be
interesting to explicitly do that for elliptically fibered
fourfolds. Since these computations allow to determine subleading
corrections and can be evaluated for various flux choices, we are
confident that they will allow to perform a solid analysis of
tunneling probabilities between various flux branches.

It is important to end with a cautionary remark on the actual
realization of an inflationary model in the proposed way. In order to
exploit the fact that there is an axion at a special point in moduli
space crucially requires to stabilize all moduli such that the
effective action is actually evaluated near this point with the axion
being the lightest field. This is a notoriously hard problem, since
often all fields start to mix in both the K\"ahler potential and
superpotential. In the K\"ahler potential, for instance, the complex
structure moduli mix with brane moduli and can correct the $\cN=1$
coordinates for the cycle volumes \cite{Jockers:2004yj,Grimm:2010ks}.
The actual field with an approximate shift-symmetry in the K\"ahler
potential will thus eventually depend on all mixings.  Furthermore,
the prefactors of instanton superpotentials used to stabilize K\"ahler
volume moduli will generically depend on the complex structure
moduli. This can ruin flatness, or at least make a separate
consideration of complex structure and K\"ahler volume moduli
questionable. Axions might be useful to identify a candidate inflaton,
but there is a good chance that it is not harder to build an
inflationary model by simply looking for a flat direction in a scalar
potential derived from string theory. In the end the hardest task is
to determine in a robust way the effective scalar potentials that
arise in string theory.

\acknowledgments

We would like to thank Ralph Blumenhagen, Luis E. Iba\~nez, Hans
Jockers, Denis Klevers, and Diego Regalado for illuminating
discussions. I.G.-E. thanks N.~Hasegawa for kind encouragement and
constant support. I.V. is supported through the FPU grant AP-2012-2690
and would like to thank the Max Planck Institute for Physics in Munich
for their hospitality during her visit in April-June 2014, when this
project was initiated.

\bibliographystyle{JHEP}
\bibliography{refs}

\providecommand{\href}[2]{#2}\begingroup\raggedright\begin{thebibliography}{10}

\bibitem{Baumann:2014nda}
D.~Baumann and L.~McAllister, {\it {Inflation and String Theory}},
  \href{http://xxx.lanl.gov/abs/1404.2601}{{\tt arXiv:1404.2601}}.

\bibitem{Douglas:2006es}
M.~R. Douglas and S.~Kachru, {\it {Flux compactification}},  {\em
  Rev.Mod.Phys.} {\bf 79} (2007) 733--796,
  [\href{http://xxx.lanl.gov/abs/hep-th/0610102}{{\tt hep-th/0610102}}].

\bibitem{Blumenhagen:2006ci}
R.~Blumenhagen, B.~Kors, D.~Lust, and S.~Stieberger, {\it {Four-dimensional
  String Compactifications with D-Branes, Orientifolds and Fluxes}},  {\em
  Phys.Rept.} {\bf 445} (2007) 1--193,
  [\href{http://xxx.lanl.gov/abs/hep-th/0610327}{{\tt hep-th/0610327}}].

\bibitem{Ade:2014xna}
{\bf BICEP2 Collaboration} Collaboration, P.~Ade {\em et.~al.}, {\it {Detection
  of B-Mode Polarization at Degree Angular Scales by BICEP2}},  {\em
  Phys.Rev.Lett.} {\bf 112} (2014) 241101,
  [\href{http://xxx.lanl.gov/abs/1403.3985}{{\tt arXiv:1403.3985}}].

\bibitem{Adam:2014bub}
{\bf Planck Collaboration} Collaboration, R.~Adam {\em et.~al.}, {\it {Planck
  intermediate results. XXX. The angular power spectrum of polarized dust
  emission at intermediate and high Galactic latitudes}},
  \href{http://xxx.lanl.gov/abs/1409.5738}{{\tt arXiv:1409.5738}}.

\bibitem{Kim:2004rp}
J.~E. Kim, H.~P. Nilles, and M.~Peloso, {\it {Completing natural inflation}},
  {\em JCAP} {\bf 0501} (2005) 005,
  [\href{http://xxx.lanl.gov/abs/hep-ph/0409138}{{\tt hep-ph/0409138}}].

\bibitem{Kappl:2014lra}
R.~Kappl, S.~Krippendorf, and H.~P. Nilles, {\it {Aligned Natural Inflation:
  Monodromies of two Axions}},  {\em Phys.Lett.} {\bf B737} (2014) 124--128,
  [\href{http://xxx.lanl.gov/abs/1404.7127}{{\tt arXiv:1404.7127}}].

\bibitem{Long:2014dta}
C.~Long, L.~McAllister, and P.~McGuirk, {\it {Aligned Natural Inflation in
  String Theory}},  {\em Phys.Rev.} {\bf D90} (2014) 023501,
  [\href{http://xxx.lanl.gov/abs/1404.7852}{{\tt arXiv:1404.7852}}].

\bibitem{Gao:2014uha}
X.~Gao, T.~Li, and P.~Shukla, {\it {Combining Universal and Odd RR Axions for
  Aligned Natural Inflation}},  {\em JCAP} {\bf 1410} (2014), no.~10 048,
  [\href{http://xxx.lanl.gov/abs/1406.0341}{{\tt arXiv:1406.0341}}].

\bibitem{Liddle:1998jc}
A.~R. Liddle, A.~Mazumdar, and F.~E. Schunck, {\it {Assisted inflation}},  {\em
  Phys.Rev.} {\bf D58} (1998) 061301,
  [\href{http://xxx.lanl.gov/abs/astro-ph/9804177}{{\tt astro-ph/9804177}}].

\bibitem{Dimopoulos:2005ac}
S.~Dimopoulos, S.~Kachru, J.~McGreevy, and J.~G. Wacker, {\it {N-flation}},
  {\em JCAP} {\bf 0808} (2008) 003,
  [\href{http://xxx.lanl.gov/abs/hep-th/0507205}{{\tt hep-th/0507205}}].

\bibitem{Grimm:2007hs}
T.~W. Grimm, {\it {Axion inflation in type II string theory}},  {\em Phys.Rev.}
  {\bf D77} (2008) 126007, [\href{http://xxx.lanl.gov/abs/0710.3883}{{\tt
  arXiv:0710.3883}}].

\bibitem{Cicoli:2014sva}
M.~Cicoli, K.~Dutta, and A.~Maharana, {\it {N-flation with Hierarchically Light
  Axions in String Compactifications}},  {\em JCAP} {\bf 1408} (2014) 012,
  [\href{http://xxx.lanl.gov/abs/1401.2579}{{\tt arXiv:1401.2579}}].

\bibitem{Bachlechner:2014hsa}
T.~C. Bachlechner, M.~Dias, J.~Frazer, and L.~McAllister, {\it {A New Angle on
  Chaotic Inflation}},  \href{http://xxx.lanl.gov/abs/1404.7496}{{\tt
  arXiv:1404.7496}}.

\bibitem{Silverstein:2008sg}
E.~Silverstein and A.~Westphal, {\it {Monodromy in the CMB: Gravity Waves and
  String Inflation}},  {\em Phys.Rev.} {\bf D78} (2008) 106003,
  [\href{http://xxx.lanl.gov/abs/0803.3085}{{\tt arXiv:0803.3085}}].

\bibitem{McAllister:2008hb}
L.~McAllister, E.~Silverstein, and A.~Westphal, {\it {Gravity Waves and Linear
  Inflation from Axion Monodromy}},  {\em Phys.Rev.} {\bf D82} (2010) 046003,
  [\href{http://xxx.lanl.gov/abs/0808.0706}{{\tt arXiv:0808.0706}}].

\bibitem{Kaloper:2008fb}
N.~Kaloper and L.~Sorbo, {\it {A Natural Framework for Chaotic Inflation}},
  {\em Phys.Rev.Lett.} {\bf 102} (2009) 121301,
  [\href{http://xxx.lanl.gov/abs/0811.1989}{{\tt arXiv:0811.1989}}].

\bibitem{Palti:2014kza}
E.~Palti and T.~Weigand, {\it {Towards large r from [p, q]-inflation}},  {\em
  JHEP} {\bf 1404} (2014) 155, [\href{http://xxx.lanl.gov/abs/1403.7507}{{\tt
  arXiv:1403.7507}}].

\bibitem{Marchesano:2014mla}
F.~Marchesano, G.~Shiu, and A.~M. Uranga, {\it {F-term Axion Monodromy
  Inflation}},  {\em JHEP} {\bf 1409} (2014) 184,
  [\href{http://xxx.lanl.gov/abs/1404.3040}{{\tt arXiv:1404.3040}}].

\bibitem{Hebecker:2014eua}
A.~Hebecker, S.~C. Kraus, and L.~T. Witkowski, {\it {D7-Brane Chaotic
  Inflation}},  {\em Phys.Lett.} {\bf B737} (2014) 16--22,
  [\href{http://xxx.lanl.gov/abs/1404.3711}{{\tt arXiv:1404.3711}}].

\bibitem{Blumenhagen:2014gta}
R.~Blumenhagen and E.~Plauschinn, {\it {Towards Universal Axion Inflation and
  Reheating in String Theory}},  {\em Phys.Lett.} {\bf B736} (2014) 482--487,
  [\href{http://xxx.lanl.gov/abs/1404.3542}{{\tt arXiv:1404.3542}}].

\bibitem{Ibanez:2014kia}
L.~E. Ibáñez and I.~Valenzuela, {\it {The inflaton as an MSSM Higgs and open
  string modulus monodromy inflation}},  {\em Phys.Lett.} {\bf B736} (2014)
  226--230, [\href{http://xxx.lanl.gov/abs/1404.5235}{{\tt arXiv:1404.5235}}].

\bibitem{Arends:2014qca}
M.~Arends, A.~Hebecker, K.~Heimpel, S.~C. Kraus, D.~Lust, {\em et.~al.}, {\it
  {D7-Brane Moduli Space in Axion Monodromy and Fluxbrane Inflation}},  {\em
  Fortsch.Phys.} {\bf 62} (2014) 647--702,
  [\href{http://xxx.lanl.gov/abs/1405.0283}{{\tt arXiv:1405.0283}}].

\bibitem{McAllister:2014mpa}
L.~McAllister, E.~Silverstein, A.~Westphal, and T.~Wrase, {\it {The Powers of
  Monodromy}},  {\em JHEP} {\bf 1409} (2014) 123,
  [\href{http://xxx.lanl.gov/abs/1405.3652}{{\tt arXiv:1405.3652}}].

\bibitem{Franco:2014hsa}
S.~Franco, D.~Galloni, A.~Retolaza, and A.~Uranga, {\it {Axion Monodromy
  Inflation on Warped Throats}},  \href{http://xxx.lanl.gov/abs/1405.7044}{{\tt
  arXiv:1405.7044}}.

\bibitem{Ibanez:2014swa}
L.~E. Ibanez, F.~Marchesano, and I.~Valenzuela, {\it {Higgs-otic Inflation and
  String Theory}},  \href{http://xxx.lanl.gov/abs/1411.5380}{{\tt
  arXiv:1411.5380}}.

\bibitem{Grimm:2005fa}
T.~W. Grimm, {\it {The Effective action of type II Calabi-Yau orientifolds}},
  {\em Fortsch.Phys.} {\bf 53} (2005) 1179--1271,
  [\href{http://xxx.lanl.gov/abs/hep-th/0507153}{{\tt hep-th/0507153}}].

\bibitem{Vafa:1996xn}
C.~Vafa, {\it {Evidence for F theory}},  {\em Nucl.Phys.} {\bf B469} (1996)
  403--418, [\href{http://xxx.lanl.gov/abs/hep-th/9602022}{{\tt
  hep-th/9602022}}].

\bibitem{Denef:2008wq}
F.~Denef, {\it {Les Houches Lectures on Constructing String Vacua}},
  \href{http://xxx.lanl.gov/abs/0803.1194}{{\tt arXiv:0803.1194}}.

\bibitem{Blumenhagen:2014nba}
R.~Blumenhagen, D.~Herschmann, and E.~Plauschinn, {\it {The Challenge of
  Realizing F-term Axion Monodromy Inflation in String Theory}},
  \href{http://xxx.lanl.gov/abs/1409.7075}{{\tt arXiv:1409.7075}}.

\bibitem{Hayashi:2014aua}
H.~Hayashi, R.~Matsuda, and T.~Watari, {\it {Issues in Complex Structure Moduli
  Inflation}},  \href{http://xxx.lanl.gov/abs/1410.7522}{{\tt
  arXiv:1410.7522}}.

\bibitem{Hebecker:2014kva}
A.~Hebecker, P.~Mangat, F.~Rompineve, and L.~T. Witkowski, {\it {Tuning and
  Backreaction in F-term Axion Monodromy Inflation}},
  \href{http://xxx.lanl.gov/abs/1411.2032}{{\tt arXiv:1411.2032}}.

\bibitem{Gukov:1999ya}
S.~Gukov, C.~Vafa, and E.~Witten, {\it {CFT's from Calabi-Yau four folds}},
  {\em Nucl.Phys.} {\bf B584} (2000) 69--108,
  [\href{http://xxx.lanl.gov/abs/hep-th/9906070}{{\tt hep-th/9906070}}].

\bibitem{Giddings:2001yu}
S.~B. Giddings, S.~Kachru, and J.~Polchinski, {\it {Hierarchies from fluxes in
  string compactifications}},  {\em Phys.Rev.} {\bf D66} (2002) 106006,
  [\href{http://xxx.lanl.gov/abs/hep-th/0105097}{{\tt hep-th/0105097}}].

\bibitem{Cox:2000vi}
D.~Cox and S.~Katz, {\it {Mirror symmetry and algebraic geometry}}, .

\bibitem{Hori:2003ic}
K.~Hori, S.~Katz, A.~Klemm, R.~Pandharipande, R.~Thomas, {\em et.~al.}, {\it
  {Mirror symmetry}}, .

\bibitem{Aspinwall:2004jr}
P.~S. Aspinwall, {\it {D-branes on Calabi-Yau manifolds}},
  \href{http://xxx.lanl.gov/abs/hep-th/0403166}{{\tt hep-th/0403166}}.

\bibitem{Berglund:1993ax}
P.~Berglund, P.~Candelas, X.~De~La~Ossa, A.~Font, T.~Hubsch, {\em et.~al.},
  {\it {Periods for Calabi-Yau and Landau-Ginzburg vacua}},  {\em Nucl.Phys.}
  {\bf B419} (1994) 352--403,
  [\href{http://xxx.lanl.gov/abs/hep-th/9308005}{{\tt hep-th/9308005}}].

\bibitem{Hosono:1993qy}
S.~Hosono, A.~Klemm, S.~Theisen, and S.-T. Yau, {\it {Mirror symmetry, mirror
  map and applications to Calabi-Yau hypersurfaces}},  {\em Commun.Math.Phys.}
  {\bf 167} (1995) 301--350,
  [\href{http://xxx.lanl.gov/abs/hep-th/9308122}{{\tt hep-th/9308122}}].

\bibitem{Hosono:1994ax}
S.~Hosono, A.~Klemm, S.~Theisen, and S.-T. Yau, {\it {Mirror symmetry, mirror
  map and applications to complete intersection Calabi-Yau spaces}},  {\em
  Nucl.Phys.} {\bf B433} (1995) 501--554,
  [\href{http://xxx.lanl.gov/abs/hep-th/9406055}{{\tt hep-th/9406055}}].

\bibitem{Greene:2000ci}
B.~R. Greene and C.~Lazaroiu, {\it {Collapsing D-branes in Calabi-Yau moduli
  space. 1.}},  {\em Nucl.Phys.} {\bf B604} (2001) 181--255,
  [\href{http://xxx.lanl.gov/abs/hep-th/0001025}{{\tt hep-th/0001025}}].

\bibitem{Mayr:2000as}
P.~Mayr, {\it {Phases of supersymmetric D-branes on Kahler manifolds and the
  McKay correspondence}},  {\em JHEP} {\bf 0101} (2001) 018,
  [\href{http://xxx.lanl.gov/abs/hep-th/0010223}{{\tt hep-th/0010223}}].

\bibitem{Grimm:2004uq}
T.~W. Grimm and J.~Louis, {\it {The Effective action of N = 1 Calabi-Yau
  orientifolds}},  {\em Nucl.Phys.} {\bf B699} (2004) 387--426,
  [\href{http://xxx.lanl.gov/abs/hep-th/0403067}{{\tt hep-th/0403067}}].

\bibitem{Benini:2012ui}
F.~Benini and S.~Cremonesi, {\it {Partition functions of N=(2,2) gauge theories
  on $S^2$ and vortices}},  \href{http://xxx.lanl.gov/abs/1206.2356}{{\tt
  arXiv:1206.2356}}.

\bibitem{Doroud:2012xw}
N.~Doroud, J.~Gomis, B.~Le~Floch, and S.~Lee, {\it {Exact Results in D=2
  Supersymmetric Gauge Theories}},  {\em JHEP} {\bf 1305} (2013) 093,
  [\href{http://xxx.lanl.gov/abs/1206.2606}{{\tt arXiv:1206.2606}}].

\bibitem{Jockers:2012dk}
H.~Jockers, V.~Kumar, J.~M. Lapan, D.~R. Morrison, and M.~Romo, {\it
  {Two-Sphere Partition Functions and Gromov-Witten Invariants}},  {\em
  Commun.Math.Phys.} {\bf 325} (2014) 1139--1170,
  [\href{http://xxx.lanl.gov/abs/1208.6244}{{\tt arXiv:1208.6244}}].

\bibitem{Halverson:2013qca}
J.~Halverson, H.~Jockers, J.~M. Lapan, and D.~R. Morrison, {\it {Perturbative
  Corrections to Kahler Moduli Spaces}},
  \href{http://xxx.lanl.gov/abs/1308.2157}{{\tt arXiv:1308.2157}}.

\bibitem{Hori:2013ika}
K.~Hori and M.~Romo, {\it {Exact Results In Two-Dimensional (2,2)
  Supersymmetric Gauge Theories With Boundary}},
  \href{http://xxx.lanl.gov/abs/1308.2438}{{\tt arXiv:1308.2438}}.

\bibitem{Kim:2013ola}
H.~Kim, S.~Lee, and P.~Yi, {\it {Exact partition functions on $\mathbb{RP}^2$
  and orientifolds}},  {\em JHEP} {\bf 1402} (2014) 103,
  [\href{http://xxx.lanl.gov/abs/1310.4505}{{\tt arXiv:1310.4505}}].

\bibitem{Freese:1990rb}
K.~Freese, J.~A. Frieman, and A.~V. Olinto, {\it {Natural inflation with pseudo
  - Nambu-Goldstone bosons}},  {\em Phys.Rev.Lett.} {\bf 65} (1990) 3233--3236.

\bibitem{Adams:1992bn}
F.~C. Adams, J.~R. Bond, K.~Freese, J.~A. Frieman, and A.~V. Olinto, {\it
  {Natural inflation: Particle physics models, power law spectra for large
  scale structure, and constraints from COBE}},  {\em Phys.Rev.} {\bf D47}
  (1993) 426--455, [\href{http://xxx.lanl.gov/abs/hep-ph/9207245}{{\tt
  hep-ph/9207245}}].

\bibitem{Vafa:2000wi}
C.~Vafa, {\it {Superstrings and topological strings at large N}},  {\em
  J.Math.Phys.} {\bf 42} (2001) 2798--2817,
  [\href{http://xxx.lanl.gov/abs/hep-th/0008142}{{\tt hep-th/0008142}}].

\bibitem{Cachazo:2001jy}
F.~Cachazo, K.~A. Intriligator, and C.~Vafa, {\it {A Large N duality via a
  geometric transition}},  {\em Nucl.Phys.} {\bf B603} (2001) 3--41,
  [\href{http://xxx.lanl.gov/abs/hep-th/0103067}{{\tt hep-th/0103067}}].

\bibitem{Dijkgraaf:2002dh}
R.~Dijkgraaf and C.~Vafa, {\it {A Perturbative window into nonperturbative
  physics}},  \href{http://xxx.lanl.gov/abs/hep-th/0208048}{{\tt
  hep-th/0208048}}.

\bibitem{Heckman:2007ub}
J.~J. Heckman and C.~Vafa, {\it {Geometrically Induced Phase Transitions at
  Large N}},  {\em JHEP} {\bf 0804} (2008) 052,
  [\href{http://xxx.lanl.gov/abs/0707.4011}{{\tt arXiv:0707.4011}}].

\bibitem{Aganagic:2007py}
M.~Aganagic, C.~Beem, and S.~Kachru, {\it {Geometric transitions and dynamical
  SUSY breaking}},  {\em Nucl.Phys.} {\bf B796} (2008) 1--24,
  [\href{http://xxx.lanl.gov/abs/0709.4277}{{\tt arXiv:0709.4277}}].

\bibitem{Grimm:2010ks}
T.~W. Grimm, {\it {The N=1 effective action of F-theory compactifications}},
  {\em Nucl.Phys.} {\bf B845} (2011) 48--92,
  [\href{http://xxx.lanl.gov/abs/1008.4133}{{\tt arXiv:1008.4133}}].

\bibitem{Alim:2012ss}
M.~Alim and E.~Scheidegger, {\it {Topological Strings on Elliptic Fibrations}},
   \href{http://xxx.lanl.gov/abs/1205.1784}{{\tt arXiv:1205.1784}}.

\bibitem{Klemm:2012sx}
A.~Klemm, J.~Manschot, and T.~Wotschke, {\it {Quantum geometry of elliptic
  Calabi-Yau manifolds}},  \href{http://xxx.lanl.gov/abs/1205.1795}{{\tt
  arXiv:1205.1795}}.

\bibitem{Grimm:2010ez}
T.~W. Grimm and T.~Weigand, {\it {On Abelian Gauge Symmetries and Proton Decay
  in Global F-theory GUTs}},  {\em Phys.Rev.} {\bf D82} (2010) 086009,
  [\href{http://xxx.lanl.gov/abs/1006.0226}{{\tt arXiv:1006.0226}}].

\bibitem{Morrison:2012ei}
D.~R. Morrison and D.~S. Park, {\it {F-Theory and the Mordell-Weil Group of
  Elliptically-Fibered Calabi-Yau Threefolds}},  {\em JHEP} {\bf 1210} (2012)
  128, [\href{http://xxx.lanl.gov/abs/1208.2695}{{\tt arXiv:1208.2695}}].

\bibitem{Braun:2013yti}
V.~Braun, T.~W. Grimm, and J.~Keitel, {\it {New Global F-theory GUTs with U(1)
  symmetries}},  {\em JHEP} {\bf 1309} (2013) 154,
  [\href{http://xxx.lanl.gov/abs/1302.1854}{{\tt arXiv:1302.1854}}].

\bibitem{Cvetic:2013nia}
M.~Cvetic, D.~Klevers, and H.~Piragua, {\it {F-Theory Compactifications with
  Multiple U(1)-Factors: Constructing Elliptic Fibrations with Rational
  Sections}},  {\em JHEP} {\bf 1306} (2013) 067,
  [\href{http://xxx.lanl.gov/abs/1303.6970}{{\tt arXiv:1303.6970}}].

\bibitem{Borchmann:2013jwa}
J.~Borchmann, C.~Mayrhofer, E.~Palti, and T.~Weigand, {\it {Elliptic fibrations
  for $SU(5)\times U(1)\times U(1)$ F-theory vacua}},  {\em Phys.Rev.} {\bf
  D88} (2013), no.~4 046005, [\href{http://xxx.lanl.gov/abs/1303.5054}{{\tt
  arXiv:1303.5054}}].

\bibitem{Grimm:2013oga}
T.~W. Grimm, A.~Kapfer, and J.~Keitel, {\it {Effective action of 6D F-Theory
  with U(1) factors: Rational sections make Chern-Simons terms jump}},  {\em
  JHEP} {\bf 1307} (2013) 115, [\href{http://xxx.lanl.gov/abs/1305.1929}{{\tt
  arXiv:1305.1929}}].

\bibitem{Braun:2013nqa}
V.~Braun, T.~W. Grimm, and J.~Keitel, {\it {Geometric Engineering in Toric
  F-Theory and GUTs with U(1) Gauge Factors}},  {\em JHEP} {\bf 1312} (2013)
  069, [\href{http://xxx.lanl.gov/abs/1306.0577}{{\tt arXiv:1306.0577}}].

\bibitem{Borchmann:2013hta}
J.~Borchmann, C.~Mayrhofer, E.~Palti, and T.~Weigand, {\it {SU(5) Tops with
  Multiple U(1)s in F-theory}},  {\em Nucl.Phys.} {\bf B882} (2014) 1--69,
  [\href{http://xxx.lanl.gov/abs/1307.2902}{{\tt arXiv:1307.2902}}].

\bibitem{Cvetic:2013qsa}
M.~Cvetic, D.~Klevers, H.~Piragua, and P.~Song, {\it {Elliptic fibrations with
  rank three Mordell-Weil group: F-theory with U(1) x U(1) x U(1) gauge
  symmetry}},  {\em JHEP} {\bf 1403} (2014) 021,
  [\href{http://xxx.lanl.gov/abs/1310.0463}{{\tt arXiv:1310.0463}}].

\bibitem{Braun:2014oya}
V.~Braun and D.~R. Morrison, {\it {F-theory on Genus-One Fibrations}},
  \href{http://xxx.lanl.gov/abs/1401.7844}{{\tt arXiv:1401.7844}}.

\bibitem{Morrison:2014era}
D.~R. Morrison and W.~Taylor, {\it {Sections, multisections, and U(1) fields in
  F-theory}},  \href{http://xxx.lanl.gov/abs/1404.1527}{{\tt arXiv:1404.1527}}.

\bibitem{Kuntzler:2014ila}
M.~Kuntzler and S.~Schafer-Nameki, {\it {Tate Trees for Elliptic Fibrations
  with Rank one Mordell-Weil group}},
  \href{http://xxx.lanl.gov/abs/1406.5174}{{\tt arXiv:1406.5174}}.

\bibitem{Klevers:2014bqa}
D.~Klevers, D.~K.~M. Pena, P.-K. Oehlmann, H.~Piragua, and J.~Reuter, {\it
  {F-Theory on all Toric Hypersurface Fibrations and its Higgs Branches}},
  \href{http://xxx.lanl.gov/abs/1408.4808}{{\tt arXiv:1408.4808}}.

\bibitem{Braun:2014qka}
V.~Braun, T.~W. Grimm, and J.~Keitel, {\it {Complete Intersection Fibers in
  F-Theory}},  \href{http://xxx.lanl.gov/abs/1411.2615}{{\tt arXiv:1411.2615}}.

\bibitem{Lawrie:2014uya}
C.~Lawrie and D.~Sacco, {\it {Tate's Algorithm for F-theory GUTs with two
  U(1)s}},  \href{http://xxx.lanl.gov/abs/1412.4125}{{\tt arXiv:1412.4125}}.

\bibitem{Anderson:2014yva}
L.~B. Anderson, I.~García-Etxebarria, T.~W. Grimm, and J.~Keitel, {\it
  {Physics of F-theory compactifications without section}},
  \href{http://xxx.lanl.gov/abs/1406.5180}{{\tt arXiv:1406.5180}}.

\bibitem{Garcia-Etxebarria:2014qua}
I.~García-Etxebarria, T.~W. Grimm, and J.~Keitel, {\it {Yukawas and discrete
  symmetries in F-theory compactifications without section}},  {\em JHEP} {\bf
  1411} (2014) 125, [\href{http://xxx.lanl.gov/abs/1408.6448}{{\tt
  arXiv:1408.6448}}].

\bibitem{Mayrhofer:2014haa}
C.~Mayrhofer, E.~Palti, O.~Till, and T.~Weigand, {\it {Discrete Gauge
  Symmetries by Higgsing in four-dimensional F-Theory Compactifications}},
  \href{http://xxx.lanl.gov/abs/1408.6831}{{\tt arXiv:1408.6831}}.

\bibitem{Mayrhofer:2014laa}
C.~Mayrhofer, E.~Palti, O.~Till, and T.~Weigand, {\it {On Discrete Symmetries
  and Torsion Homology in F-Theory}},
  \href{http://xxx.lanl.gov/abs/1410.7814}{{\tt arXiv:1410.7814}}.

\bibitem{Mayr:1996sh}
P.~Mayr, {\it {Mirror symmetry, N=1 superpotentials and tensionless strings on
  Calabi-Yau four folds}},  {\em Nucl.Phys.} {\bf B494} (1997) 489--545,
  [\href{http://xxx.lanl.gov/abs/hep-th/9610162}{{\tt hep-th/9610162}}].

\bibitem{Klemm:1996ts}
A.~Klemm, B.~Lian, S.~Roan, and S.-T. Yau, {\it {Calabi-Yau fourfolds for M
  theory and F theory compactifications}},  {\em Nucl.Phys.} {\bf B518} (1998)
  515--574, [\href{http://xxx.lanl.gov/abs/hep-th/9701023}{{\tt
  hep-th/9701023}}].

\bibitem{Alim:2009bx}
M.~Alim, M.~Hecht, H.~Jockers, P.~Mayr, A.~Mertens, {\em et.~al.}, {\it {Hints
  for Off-Shell Mirror Symmetry in type II/F-theory Compactifications}},  {\em
  Nucl.Phys.} {\bf B841} (2010) 303--338,
  [\href{http://xxx.lanl.gov/abs/0909.1842}{{\tt arXiv:0909.1842}}].

\bibitem{Grimm:2009ef}
T.~W. Grimm, T.-W. Ha, A.~Klemm, and D.~Klevers, {\it {Computing Brane and Flux
  Superpotentials in F-theory Compactifications}},  {\em JHEP} {\bf 1004}
  (2010) 015, [\href{http://xxx.lanl.gov/abs/0909.2025}{{\tt
  arXiv:0909.2025}}].

\bibitem{Bizet:2014uua}
N.~C. Bizet, A.~Klemm, and D.~V. Lopes, {\it {Landscaping with fluxes and the
  E8 Yukawa Point in F-theory}},  \href{http://xxx.lanl.gov/abs/1404.7645}{{\tt
  arXiv:1404.7645}}.

\bibitem{Hebecker:2011hk}
A.~Hebecker, S.~C. Kraus, D.~Lust, S.~Steinfurt, and T.~Weigand, {\it
  {Fluxbrane Inflation}},  {\em Nucl.Phys.} {\bf B854} (2012) 509--551,
  [\href{http://xxx.lanl.gov/abs/1104.5016}{{\tt arXiv:1104.5016}}].

\bibitem{Hebecker:2012aw}
A.~Hebecker, S.~C. Kraus, M.~Kuntzler, D.~Lust, and T.~Weigand, {\it
  {Fluxbranes: Moduli Stabilisation and Inflation}},  {\em JHEP} {\bf 1301}
  (2013) 095, [\href{http://xxx.lanl.gov/abs/1207.2766}{{\tt
  arXiv:1207.2766}}].

\bibitem{Olver:2010:NHMF}
F.~W.~J. Olver, D.~W. Lozier, R.~F. Boisvert, and C.~W. Clark, eds., {\em {NIST
  Handbook of Mathematical Functions}}.
\newblock Cambridge University Press, New York, NY, 2010.
\newblock Print companion to \cite{NIST:DLMF}.

\bibitem{Libgober}
A.~{Libgober}, {\it {Chern classes and the periods of mirrors}},  {\em ArXiv
  Mathematics e-prints} (Mar., 1998)
  [\href{http://xxx.lanl.gov/abs/math/9803119}{{\tt math/9803119}}].

\bibitem{Iritani1}
H.~{Iritani}, {\it {Real and integral structures in quantum cohomology I: toric
  orbifolds}},  {\em ArXiv e-prints} (Dec., 2007)
  [\href{http://xxx.lanl.gov/abs/0712.2204}{{\tt arXiv:0712.2204}}].

\bibitem{KKP}
L.~{Katzarkov}, M.~{Kontsevich}, and T.~{Pantev}, {\it {Hodge theoretic aspects
  of mirror symmetry}},  {\em ArXiv e-prints} (May, 2008)
  [\href{http://xxx.lanl.gov/abs/0806.0107}{{\tt arXiv:0806.0107}}].

\bibitem{Iritani2}
H.~{Iritani}, {\it {An integral structure in quantum cohomology and mirror
  symmetry for toric orbifolds}},  {\em ArXiv e-prints} (Mar., 2009)
  [\href{http://xxx.lanl.gov/abs/0903.1463}{{\tt arXiv:0903.1463}}].

\bibitem{Huang:2006hq}
M.-x. Huang, A.~Klemm, and S.~Quackenbush, {\it {Topological string theory on
  compact Calabi-Yau: Modularity and boundary conditions}},  {\em Lect.Notes
  Phys.} {\bf 757} (2009) 45--102,
  [\href{http://xxx.lanl.gov/abs/hep-th/0612125}{{\tt hep-th/0612125}}].

\bibitem{ChenYangYui}
Y.-H. {Chen}, Y.~{Yang}, and N.~{Yui}, {\it {Monodromy of Picard-Fuchs
  differential equations for Calabi-Yau threefolds}},  {\em ArXiv Mathematics
  e-prints} (May, 2006) [\href{http://xxx.lanl.gov/abs/math/0605675}{{\tt
  math/0605675}}].

\bibitem{Candelas:1990rm}
P.~Candelas, X.~C. De~La~Ossa, P.~S. Green, and L.~Parkes, {\it {A Pair of
  Calabi-Yau manifolds as an exactly soluble superconformal theory}},  {\em
  Nucl.Phys.} {\bf B359} (1991) 21--74.

\bibitem{Andreas:1999ty}
B.~Andreas, G.~Curio, and A.~Klemm, {\it {Towards the Standard Model spectrum
  from elliptic Calabi-Yau}},  {\em Int.J.Mod.Phys.} {\bf A19} (2004) 1987,
  [\href{http://xxx.lanl.gov/abs/hep-th/9903052}{{\tt hep-th/9903052}}].

\bibitem{Intriligator:2012ue}
K.~Intriligator, H.~Jockers, P.~Mayr, D.~R. Morrison, and M.~R. Plesser, {\it
  {Conifold Transitions in M-theory on Calabi-Yau Fourfolds with Background
  Fluxes}},  {\em Adv.Theor.Math.Phys.} {\bf 17} (2013) 601--699,
  [\href{http://xxx.lanl.gov/abs/1203.6662}{{\tt arXiv:1203.6662}}].

\bibitem{Grimm:2014vva}
T.~W. Grimm, {\it {Axion Inflation in F-theory}},
  \href{http://xxx.lanl.gov/abs/1404.4268}{{\tt arXiv:1404.4268}}.

\bibitem{Jockers:2004yj}
H.~Jockers and J.~Louis, {\it {The Effective action of D7-branes in N = 1
  Calabi-Yau orientifolds}},  {\em Nucl.Phys.} {\bf B705} (2005) 167--211,
  [\href{http://xxx.lanl.gov/abs/hep-th/0409098}{{\tt hep-th/0409098}}].

\bibitem{NIST:DLMF}
``{NIST Digital Library of Mathematical Functions}.''
  \url{http://dlmf.nist.gov/}, Release 1.0.9 of 2014-08-29.
\newblock Online companion to \cite{Olver:2010:NHMF}.

\end{thebibliography}\endgroup

\end{document}